\documentclass[nolinenumbers]{aa}  

\usepackage{graphicx}
\usepackage{txfonts}
\usepackage{svg}
\usepackage{booktabs}
\usepackage{lipsum}
\usepackage{subcaption}         % necessary for continued figures, example in section 3
\usepackage{lscape}             % to rotate a single page table, example in appendix.
\usepackage{placeins}           % useful with \FloatBarrier, to keep 
\usepackage{amsmath}
\usepackage{amsfonts}
\usepackage{hyperref}
\hypersetup{
    colorlinks=true,
    linkcolor=blue,
    citecolor=blue,
    urlcolor=blue
}
\usepackage{orcidlink}
\usepackage{color}
\begin{document}
   \title{The PAU Survey: Self-supervised image denoising for flux-preserving photometry}

   %\subtitle{Subtitle}

%%%%%%%%%%%%%%%%%%%%%%%%%%%%%%%%%%%%%%%%
% Please do not include ORCIDs next to author names.
% Only ORCIDs authenticated by individual authors in EDP Sciences editorial system will be taken into account.
% ORCIDs included here will be removed.
%%%%%%%%%%%%%%%%%%%%%%%%%%%%%%%%%%%%%%%%

   \author{J.~Chen\orcidlink{0000-0003-3877-6579}\inst{1,2}\fnmsep\thanks{jchan@pic.es}
        \and M.~Eriksen\orcidlink{0000-0003-0601-0990}\inst{1,2}\fnmsep\thanks{eriksen@pic.es}
        \and D.~Navarro-Gironés\orcidlink{0000-0003-0507-372X}\inst{3}
        \and P.~Renard\orcidlink{0000-0002-5953-4491}\inst{4,5}
        \and E.~Gaztanaga\orcidlink{0000-0001-9632-0815}\inst{6,4,5}
        \and J.~Carretero\orcidlink{0000-0002-3130-0204}\inst{7,2}
        \and R.~Casas\orcidlink{0000-0002-8165-5601}\inst{4,5}
        \and F.~J.~Castander\orcidlink{0000-0001-7316-4573}\inst{4,5}
        \and E.~Fernandez\orcidlink{0000-0002-6405-9488}\inst{1}
        \and J.~Garc\'ia-Bellido\orcidlink{0000-0002-9370-8360}\inst{8}
        \and R.~Miquel\orcidlink{0000-0002-6610-4836}\inst{1,9}
        \and C.~Padilla\orcidlink{0000-0001-7951-0166}\inst{1}
        \and E.~Sánchez\orcidlink{0000-0002-9646-8198}\inst{7}
        \and I.~Sevilla-Noarbe\orcidlink{0000-0002-1831-1953}\inst{7}
        \and P.~Tallada-Cresp\'{i}\orcidlink{0000-0002-1336-8328}\inst{7,2}
        }

\institute{Institut de F\'{i}sica d’Altes Energies (IFAE), The Barcelona Institute of Science and Technology, Campus UAB, 08193 Bellaterra (Barcelona), Spain\\
        \and Port d'Informaci\'{o} Cient\'{i}fica (PIC), Campus UAB, C. Albareda s/n, 08193 Bellaterra (Barcelona), Spain\\
        \and Leiden Observatory, Leiden University, Einsteinweg 55, 2333 CC Leiden, the Netherlands\\
        \and Institute of Space Sciences (ICE, CSIC), Campus UAB, Carrer de Can Magrans, s/n, 08193 Barcelona, Spain\\
        \and Institut d'Estudis Espacials de Catalunya (IEEC), E-08860 Castelldefels (Barcelona), Spain\\
        \and Institute of Cosmology and Gravitation, University of Portsmouth, Portsmouth, UK \\
        \and Centro de Investigaciones Energéticas, Medioambientales y Tecnológicas (CIEMAT), Avenida Complutense 40, 28040 Madrid, Spain\\
        \and Instituto de F\'isica Te\'orica CSIC/UAM, Universidad Aut\'onoma de Madrid, 28049 Madrid, Spain \\
        \and Instituci\'o Catalana de Recerca i Estudis Avan\c{c}ats (ICREA), Passeig de Llu\'{\i}s Companys, 23, 08010 Barcelona, Spain 
        }
\date{Received: date / Accepted: date}

% \abstract{}{}{}{}{}
% 5 {} token are mandatory
 
  \abstract
   {Deep photometric surveys increasingly rely on low signal-to-noise imaging to detect and characterise faint galaxies. In this regime, noise degrades image quality and affects source detection, flux measurements, photometric-redshift estimation, and sample selection. Image denoising methods can therefore improve the scientific exploitation of these data, provided that they preserve the photometric information encoded in the images. Self‑supervised denoising methods are particularly attractive because they do not require ground‑truth images, but extending them to quantitative astronomical photometry requires controlling the systematic flux biases they may introduce.}
   {We evaluate and improve Noise2Void (N2V) for denoising real astronomical images, focusing on preserving fluxes,  reducing photometric scatter, and  retaining the source main features across a wide magnitude range.} 
   {We use real observations from the Physics of the Accelerating Universe Survey (PAUS) together with PAUS-like simulated images. We first assess the standard N2V method and show that it systematically underestimates galaxy fluxes, especially for faint galaxies. To mitigate this bias, we modify the N2V architecture with a bounded-output layer that limits large negative outputs. 
   We then fine-tune the model with low-rank adaptation (LoRA) and two flux-aware loss terms designed to preserve the flux scale and improve repeated-exposure flux consistency.}
    {Standard N2V improves the visual appearance of the images, but introduces strong flux biases for faint galaxies. For PAUS galaxies with $22 \leq i_{\mathrm{AB}} < 23$ in the NB645 band, the standard N2V flux-ratio distribution peaks far below unity, while the bounded-output model restores the peak close to the original PAUS flux scale. The flux-aware LoRA fine-tuning further reduces the width of the repeated-exposure flux-difference distribution to $\sigma_{68}=0.529$ in NB645, corresponding to an effective 47\% reduction in random flux scatter relative to the original PAUS uncertainty scale, while keeping the median flux ratio close to unity.
    These results show that N2V-based denoising can reduce noise while preserving the photometric information required for quantitative astronomical analyses.}
  {Our results demonstrate that denoising methods for astronomical surveys must be evaluated with photometric diagnostics, rather than image-quality metrics alone. Flux-aware modifications to N2V reduce photometric biases and improve repeated-exposure consistency in PAUS, demonstrating that self-supervised denoising can be adapted to low-SNR survey images while preserving the flux information needed for quantitative analysis.}

   \maketitle
   \nolinenumbers

%%%%%%%%%%%%%%%%%%%%%%%%%%%%%%%%%%%%%%%%%%%%%%%%%%%%%%%%%%%%%%
\section{Introduction}
Noise is a major source of uncertainty in deep astronomical imaging. It arises from photon-counting statistics, detector readout noise \citep{1991PASP..103..122N}, and sky-background variations \citep{howell2000handbook, starck2007astronomical}, degrading both data quality and downstream measurements. In particular, these noise sources can obscure faint extended features of low-surface-brightness galaxies, hinder source separation in crowded stellar fields, and introduce biases or additional uncertainties into morphological and photometric measurements. 
These effects propagate into a wide range of quantitative astronomical analyses, including weak gravitational lensing \citep{mandelbaum2018weak, 1992ApJ...388..272K, refregier2003weak}, galaxy-clustering measurements \citep{2021MNRAS.504.4667A}, supernova light-curve analysis \citep{2018ApJ...859..101S}, cosmological-parameter inference \citep{chen2023test, 2025EPJC...85.1005C, 2023ApJS..268...67H, 2021ApJS..254...43W}, and photometric-redshift estimation \citep{2000A&A...363..476B}. Consequently, image-restoration methods can be valuable components of imaging-survey data-processing pipelines, but their scientific use requires that they preserve photometric information and be validated with diagnostics tied to the intended measurements \citep{robertson2019galaxy}.
 
 Image denoising is an ill-posed problem: suppressing noise requires assumptions about the underlying signal, and inappropriate assumptions can blur compact sources, erase faint emission, or bias flux measurements. These assumptions can be viewed as image priors, and their design has shaped the development of denoising methods (see \citealt{elad2023image} for a review). Classical methods, such as linear filtering \citep{gonzalez2009digital} and wavelet-based approaches \citep{starck2015sparse}, impose explicit image priors, such as smoothness or sparsity, together with assumptions about the noise properties.
 In astronomy, however, high-precision measurements often avoids full image restoration because restoration priors can bias the quantities of interest. Instead, many analyses use task-specific inference methods, such as model-fitting photometry \citep{2002AJ....124..266P, 1996A&AS..117..393B}, which estimate fluxes and source parameters directly from noisy data. 

Rather than imposing explicit priors, deep learning networks can learn
image priors directly from training data \citep{lecun2015deep}. Many high-performing neural denoisers are trained in a supervised manner, 
using paired noisy or degraded inputs and clean reference images, often generated synthetically \citep{2019arXiv190605478M, zhang2021plug, zhang2023practical}. Applying these methods to astronomical imaging is challenging for two main reasons. First, noiseless ground‑truth images are generally unavailable for real astronomical observations. Second, many neural denoisers are trained or pre-trained on natural-image datasets \citep{meng2020large, sriram2020end}, whose image statistics differ markedly from those of astronomical images. Astronomical sources often occupy only a small fraction of an image and can be faint relative to structured sky backgrounds, while the noise is shaped by the detector, observing conditions, and data-reduction pipeline \citep{chen2021equivariant}. 

These limitations motivate self‑supervised denoising strategies that are trained directly on the target observations, rather than on external clean reference images. A family of self-supervised approaches exploits statistical assumptions about the noise to train models from noisy data alone \citep{2017MNRAS.467L.110S, 2018Noise2Noise, batson2019noise2self}. This makes them particularly attractive for astronomical surveys, where paired noisy–clean images are generally unavailable. Although clean reference images are lacking, repeated observations, simulations, and survey-specific noise estimates can be used to test the assumptions of these methods and to validate their impact on measured fluxes.

In this work, we apply Noise2Void (N2V; \citealt{krull2019noise2void}) to narrow-band images from 
the Physics of the Accelerating Universe Survey (PAUS). N2V is a self-supervised denoising method that can be trained directly on noisy images. We show that visually cleaner denoised images can yield biased photometry: standard N2V systematically underestimate the fluxes of faint galaxies, both in simulations and in PAUS observations. In the low-SNR regime, local noise fluctuations can be comparable to the galaxy signal, making flux preservation particularly challenging. This bias reflects a mismatch between the standard N2V objective, which optimised pixel-level reconstruction, and the photometric quantities needed for science. 
We address this by modifying the network architecture with a bounded-output layer and by fine-tuning the model with flux-aware loss terms.

This paper is organized as follows. Section~\ref{sec:2} describes the PAUS observations, the simulated images, and the photometric diagnostics used in the analysis. Section~\ref{sec:3} presents the denoising methods, including the classical baselines, standard N2V, Probabilistic Noise2Void (PN2V; \citealt{Krull_2020}) and the proposed flux-aware modifications. In Sect.~\ref{sec:4}, we test the models on simulated data and PAUS observations, focusing on image quality, flux preservation, and repeated-exposure flux consistency. Besides, we also do tests with the denoised flux.
Finally, Sect.~\ref{sec:5} discusses the implications of the results and summarises the conclusions.

%%%%%%%%%%%%%%%%%%%%%%%%%%%%%%%%%%%%%%%%%%%%%%%%%%%%%%%%%%%%%%

\section{Data and photometric diagnostics}\label{sec:2}

\subsection{PAUS image data}
PAUS measures photometric redshifts (photo-$z$) and spectral energy distributions (SEDs) using much finer wavelength sampling than conventional broad-band photometric surveys. The observations were obtained with PAUCam, using 40 narrow-band filters covering $450$--$850\,\mathrm{nm}$ in $10\,\mathrm{nm}$ steps \citep{2019AJ....157..246P}. 
We use reduced PAUS images and photometric products from the COSMOS field, based on MEMBA forced-photometry pipeline (production version v821).
The pipeline performs astrometric and photometric calibration, background subtraction, and aperture-photometry measurements \citep{tonello2019pau, 2023MNRAS.523.3287S}. We use the calibrated image products as model inputs and the PAUS flux measurements to evaluate flux preservation and repeated-exposure consistency.

We use reduced images rather than raw exposures for both scientific and practical reasons. The reduced images have undergone the standard PAUS processing steps, including bias correction, flat-fielding, astrometric and photometric calibration and background subtraction. Applying denoising to the raw exposures would require rerunning the full survey reduction pipeline after denoising. This would be computationally expensive and operationally non-trivial. Moreover, a pipeline optimised for raw  detector images would not necessarily behave in the same way when applied to denoised inputs. By working with reduced images, we can compare flux measurements before and after denoising using the same calibrated image units, catalogue positions, apertures, and masks. The implications of using reduced images, including the weak spatial noise correlations introduced by the reduction pipeline, are discussed further in Appendix~\ref{app:noisecorr}.

Throughout this work, image intensities and aperture fluxes are expressed in calibrated analogue-to-digital units per second $\mathrm{ADU}\,\mathrm{s}^{-1}$ for both PAUS observations and simulated images.

\subsection{PAUS image simulations}

GPU memory constraints prevent the direct processing of 
full PAUCam frames. We therefore generate simulated images as $96\times96$ pixel stamps, matching the dimensions of the real PAUS cutouts used for training and evaluation.

The simulated image stamps provide a controlled dataset with known source properties and noise, allowing us to quantify flux biases and denoising performance under PAUS-like conditions. They are generated using a pipeline that combines simple parametric galaxy profiles with instrumental and observational effects, described in Appendix~\ref{app:sim}.

\subsection{Aperture flux measurements}
\label{subsec:appflux}
The PAUS narrow-band images typically have $\mathrm{SNR} < 3$ at $i_{\mathrm{AB}} = 23$, making reliable photometric measurements challenging for the faintest sources. In particular, the shapes of such low-SNR galaxies cannot be robustly estimated directly from the PAUS narrow-band images. PAUS therefore uses forced-photometry based on external COSMOS, KiDS and CFHTLenS reference catalogues \citep{2023MNRAS.523.3287S}. These provide the source positions and shape information used to define fixed apertures that enclose an estimated $62.5\%$  of the source flux.

We measure fluxes using the aperture-photometry approach adopted in the standard PAU Data Management (PAUdm) processing pipeline, making
our diagnostics directly comparable to PAUdm fluxes. For PAUS observations, image stamps are centred on the catalogue positions. The aperture, background, and mask arrays are constructed using \texttt{photutils} \citep{bradley2020astropy} and \texttt{astropy} \citep{astropy:2013,astropy:2018,astropy:2022}.
The flux and background estimation are implemented as weighted sums over each image stamp. This detail is important for LoRA-N2V training (Sect.~\ref{sec:LoRA-N2V}), as it makes the aperture-flux measurement differentiable and allows flux-based losses to be evaluated during optimisation. 

Let $I_{p}$ be the pixel value and $W_{ p}$ the aperture-weight for pixel $p$. To account for fractional pixels at the aperture boundaries, $W_{p} \in [0, 1]$ is defined as the fractional area of pixel $p$. These weights are computed with \texttt{Nayo}\footnote{\url{https://github.com/PAU-survey/nayo}}, a reimplementation of the PAUdm forced-photometry method based on \texttt{photutils}. The total aperture flux is then
\begin{equation}
f_{\rm tot} = \sum_p W_p I_p.
\end{equation}
with aperture area $A_{\rm gal} = \sum_p W_p$ in pixels. The local sky background, $b_{\mathrm{sky}}$, is estimated as the median pixel value in an annular region centred on the source, with inner and outer radii of 30 and 45 pixels \citep{2023MNRAS.523.3287S}. The background-subtracted galaxy flux is then

    \begin{equation}
        f_{\mathrm{gal}} = f_{\mathrm{tot}} - A_{\mathrm{gal}}\, b_{\mathrm{sky}}.
    \end{equation}

\subsection{Flux-repeatability diagnostics}\label{sec:df}

The flux-difference statistic, $df$, is used in PAUS to compare repeated measurements of the same source and test
whether the reported flux uncertainties match the expected scatter between duplicate exposures. Following \citep{2023MNRAS.523.3287S}, we define
\begin{equation}
    df_{\mathrm{orig}} \equiv \frac{f_1 - f_2}{\sigma}
    = \frac{f_1 - f_2}{\sqrt{\sigma_1^2 + \sigma_2^2}},
    \label{eq:df}
\end{equation}
where $f_1$ and $f_2$ are duplicate calibrated flux measurements of the same galaxy in the same filter, and $\sigma_1$ and $\sigma_2$ are their uncertainties. If the measurements have independent Gaussian errors and correctly estimated uncertainties, the $df_{\mathrm{orig}}$ distribution should follow a standard normal distribution; a derivation is given in Appendix~\ref{app:df}. Deviations from this behaviour can indicate underestimated or overestimated uncertainties, correlated noise or residual systematics.

The same diagnostic is applied to evaluate denoised images with a different convention: the numerator uses the denoised fluxes ($\tilde{f}$), while the denominator is kept fixed to the original PAUS catalogue uncertainties,

\begin{equation}
df_{\mathrm{den}} \equiv
\frac{\tilde{f}_1 - \tilde{f}_2}{\sqrt{\sigma_1^2 + \sigma_2^2}} .
\label{eq:dfden}
\end{equation}

\noindent
With this convention, the width of the $df$ distribution measures the scatter between repeated denoised flux measurements relative to the original PAUS uncertainty scale. A narrower distribution therefore indicates improved repeated-exposure flux precision.

This interpretation is approximate, since $df_{\mathrm{den}}$ is also affected by flux bias common to both repeated measurements. For example, a multiplicative suppression of both denoised fluxes would also reduce the numerator of Eq.~\ref{eq:dfden}. We therefore combine the $df_{\mathrm{den}}$ diagnostic with flux-ratio measurements to assess both repeated-exposure precision and photometric bias; see Appendix~\ref{app:df} for a more detailed discussion.

\section{Denoising models}\label{sec:3}

This section describes the denoising methods used in this work. We first introduce the classical denoising baselines, followed by the N2V and PN2V models.
We then present the two N2V-based modifications developed to improve photometric performance: the constrained-output N2V and the flux-aware LoRA-N2V fine-tuning.

\subsection{Classical denoising baselines}

As reference comparisons, we include three classical denoising baselines representing patch-based, variational, and sparse-transform approaches. These are not intended as an exhaustive benchmark, but as representative non-learning comparisons for the N2V-based models. They are evaluated only on simulated images, where noiseless ground truth images are available for quantitative image-quality metrics.

Block-matching and 3D filtering (BM3D) \citep{dabov2007image, makinen2020collaborative} is a patch-based denoising method that groups similar image patches and applies collaborative filtering, assuming white Gaussian noise. Total generalized variation (TGV) denoising \citep{bredies2010total, condat2013primal} is a higher-order extension of total-variation regularization that promotes smooth reconstructions while reducing the staircasing artefacts. Finally, WaveletDictDenoiser (Wavelet) \citep{lee2019pywavelets} denoises images by thresholding wavelet coefficients, exploiting sparse representation
in a suitable wavelet basis.

These  baselines are implemented using the open-source \texttt{DeepInverse} library \citep{tachella2025deepinverse}, which provides PyTorch implementations of imaging-restoration methods. For each baseline, we performed a coarse sweep of the noise parameter on the simulated images and selected representative settings that produced visible noise suppression without introducing excessive artefacts. This tuning was intended to provide qualitative reference comparisons rather than an exhaustive optimisation of the classical methods.

\subsection{N2V}
Noise2Void (N2V; \citealt{krull2019noise2void}) is a self-supervised image-denoising method trained only on noisy observations, without noiseless reference targets. It assumes pixel-wise independent, zero-mean noise and a
spatially correlated signal. Under these assumptions, 
neighbouring pixels can predict the signal at a masked pixel, but not its specific noise realisation. 

\begin{figure}  \centering
  \includegraphics[width=1\linewidth]{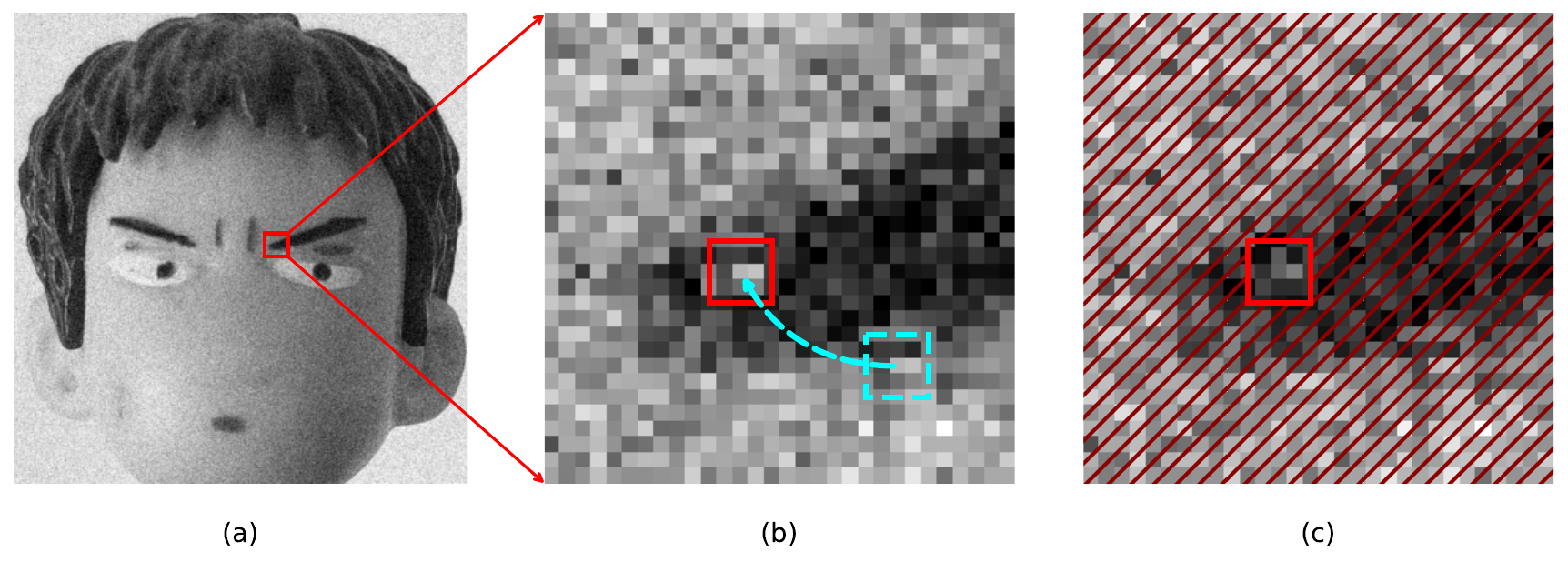}
  \caption{
  Illustration of the blind-spot masking scheme used during Noise2Void (N2V) training. 
\emph{(a)} Noisy input image. 
\emph{(b)} Extracted noisy patch, where the red square marks the target pixel and the blue square the replacement value used to mask it.
\emph{(c)} Masked input patch used for training. The hatched region is excluded from the loss, which is
evaluated only at the mask-pixel locations.}
  \label{fig:n2v_blindspot}
\end{figure}

We write the observed image as
\begin{equation}
\boldsymbol{x} = \boldsymbol{s} + \boldsymbol{n},
\end{equation}
where $\boldsymbol{x}$ is the observed image, $\boldsymbol{s}$ is the underlying signal and $\boldsymbol{n}$ is the noise. N2V uses a blind-spot strategy in which selected pixels are masked in the input and the network is trained to predict their original noisy values from the surrounding context. Figure~\ref{fig:n2v_blindspot} illustrates the N2V masking scheme. For each selected target pixel, the input value is replaced, and the loss is evaluated only at the masked pixel locations. Since the target pixel is hidden, the network cannot simply copy its noise realisation and is instead encouraged to infer the predictable signal component. 

Formally, for a masked pixel $i$, the model prediction is

\begin{equation}
\hat{s}_i = f_{\theta}\left(\hat{\boldsymbol{x}}_{\mathrm{RF}(i)}\right),
\end{equation}

\noindent
where $f_{\theta}$ is the neural network with parameters $\theta$, $\hat{\boldsymbol{x}}$ is the masked input image, and $\mathrm{RF}(i)$ denotes the receptive field around pixel $i$, excluding the target pixel. The training target is the original noisy value $x_i$, and training 
minimises a mean-squared error over the masked pixels, 

\begin{equation}\label{eq:n2vloss}
\mathcal{L}_{\mathrm{N2V}}
= \sum_j \sum_{i\, \in \, M_j}
\left[
f_{\theta}\left(\hat{\boldsymbol{x}}^{j}_{\mathrm{RF}(i)}\right)
- x_i^j
\right]^2 ,
\end{equation}

\noindent
where $j$ indexes the training images and $M_j$ is the set of masked pixel locations in image $j$. 

Our N2V implementation uses a U-Net architecture \citep{DBLP:conf/miccai/RonnebergerFB15}, an encoder--decoder network with skip connections that 
combine contextual information with fine spatial detail. Batch normalisation layers \citep{ioffe2015batch} are placed before the activation functions.

\subsection{Probabilistic Noise2Void}

Probabilistic Noise2Void (PN2V; \citealt{Krull_2020}) extends N2V by predicting a distribution for the underlying signal at each pixel, rather than a single estimate. From the blind-spot context around pixel $i$, the network predicts $p(s_i \mid \hat{x}_{\text{RF}(i)})$, which is combined with an explicit noise model $p(x_i \mid s_i)$ for the observed noisy pixel.

The predicted signal distribution is represented by $K=800$ sample channels, $\{s_i^k\}_{k=1}^K$,
which are output simultaneously by the network.
The model is trained by minimising the negative log-likelihood of the observed pixels:

\begin{equation}
 \mathcal{L}_{\text{PN2V}}= \sum_i -\log \left( \frac{1}{K} \sum_{k=1}^K p(x_i \mid s_i^k) \right).
\end{equation}

\noindent
Given the observed pixel value, the minimum mean-square-error (MMSE) estimate of the signal can be approximated as

\begin{equation}
s_i^{\text{MMSE}} \approx \frac{\sum_{k=1}^K p(x_i \mid s_i^k)\, s_i^k}{\sum_{k=1}^K p(x_i \mid s_i^k)}.
\end{equation}

Tests on PAUS images showed that the explicit PN2V noise likelihood is sensitive to weak noise correlations, 
producing background artefacts; see Appendix~\ref{app:noisecorr_impl}. We therefore do not use the full PN2V likelihood formulation in our final model. Instead, we retain the idea of repeated stochastic predictions: for each pixel,
we generate $K=800$ stochastic predictions and use their mean as the final denoised value, without PN2V likelihood reweighting.

\subsection{Constrained N2V}
\label{subsec:constrained_n2v}

The PN2V implementation described above can produce negative-value predictions in low-SNR regions. When averaged over repeated predictions, these negative values can shift the output fluxes downward and 
introduce systematic flux biases for faint sources. 
To mitigate this, we add an exponential linear unit (ELU; \citealt{2015arXiv151107289C}) after the final linear output layer, and refer to the resulting model as constrained N2V.

The ELU activation does not impose strict non-negativity, but bounds the negative output values while remaining linear for positive values. With the value used here, $\alpha=0.1$, the activation satisfies $\mathrm{ELU}(z) \geq -0.1$. Unlike ReLU \citep{Nair2010RectifiedLU}, which enforces non-negative outputs but has zero gradient for negative pre-activations, ELU retains non-zero gradients in the negative regime. We found this to give more stable training and better photometric behaviour than a strictly non-negative ReLU output. All other aspects
of the N2V training procedure are unchanged.

\subsection{LoRA-N2V}\label{sec:LoRA-N2V}
The constrained N2V model suppresses large negative predictions and thereby reduces flux loss, but it does
not exploit PAUS repeated-exposures or directly optimise flux consistency.
We therefore introduce a flux-aware fine-tuning stage based on two aperture-flux losses: a paired-difference loss, which encourages consistent fluxes for repeated observations of the same galaxy, and a photometric-bias loss, which anchors the denoised photometry to the original PAUS flux scale. Together, these losses 
reduce random flux scatter while discouraging biased
rescaling of the photometry.

\subsubsection{Paired difference loss}

The \emph{paired-difference loss} ($\mathcal{L}_{\mathrm{pd}}$) improves the repeatability of flux measurements from duplicate exposures. It uses the same repeated-exposure pair structure as the $df$ statistic defined in Sect.~\ref{sec:df}, but without normalising by the flux uncertainties. For mini-batches of $n$ matched galaxy
pairs, where each pair consists of the same galaxies in the same filter but in different exposures, we define
\begin{equation}
    \mathcal{L}_{\mathrm{pd}} = \frac{1}{n} \sum_{i=1}^{n} \left| f_i^{(1)} - f_i^{(2)} \right|,
\end{equation}
where $f_i^{(1)}$ and $f_i^{(2)}$ are the aperture fluxes measured from the two denoised exposures of the \(i\)-th galaxy pair. The paired mini-batches are constructed with a custom \texttt{PyTorch} DataLoader \citep{paszke2019pytorch}.

By itself, this loss only enforces consistency between repeated measurements. It does not prevent the network from reducing the paired difference through biased flux rescaling, motivating the photometric-bias loss introduced below.

\subsubsection{Photometric bias loss}

The \emph{photometric bias loss} ($\mathcal{L}_{\mathrm{pb}}$) prevents the network from reducing the paired-difference loss through a trivial rescaling of the fluxes. For example, multiplying all denoised fluxes by a constant below unity would reduce the scatter between repeated measurements, but bias the photometry. We therefore penalise deviations between aperture flux measured before and after denoising the
same image:

\begin{equation}
    \mathcal{L}_{\mathrm{pb}} =
    \frac{1}{N} \sum_{i=1}^{N}
    \left| f^{(i)}_{\mathrm{input}} - f^{(i)}_{\mathrm{output}} \right|,
\end{equation}
where $N$ is the number of galaxy measurements in the batch, and
$f^{(i)}_{\mathrm{input}}$ and $f^{(i)}_{\mathrm{output}}$ are the aperture fluxes before and after denoising the $i$-th image, respectively. This loss does not assume that the input flux is noiseless, but discourages systematic shifts relative to the original PAUS flux scale. The effect of such flux rescalings on the $df$ distribution is illustrated in Appendix~\ref{app:df}.

\subsubsection{Flux-aware LoRA fine-tuning}

Optimising the flux-based losses introduced above requires paired galaxy observations and aperture-flux measurement during training. Updating all N2V parameters with these losses would be 
substantially more expensive than the original N2V training. We therefore use low-rank adaptation (LoRA; \citealt{hu2022lora}) as a parameter-efficient fine-tuning strategy.

LoRA keeps the weights of the pre-trained constrained N2V backbone fixed and adds trainable low-rank matrices to selected layers. The resulting model, LoRA-N2V, can adapt to the flux-based objective while largely preserving the denoising behaviour learned by the base model. 
Although motivated by PN2V-style repeated stochastic predictions, LoRA-N2V remains N2V-based and does not include the explicit PN2V noise likelihood or noise model.
The LoRA modules are implemented using the \texttt{peft} library \citep{peft}.

A central aspect of this training scheme is that aperture-flux measurements are implemented as a differentiable \texttt{PyTorch} operation \citep{paszke2019pytorch}. The model first produces denoised images, after which aperture fluxes are measured from the input and denoised images using the aperture definitions described in Sect.~\ref{subsec:appflux}. These fluxes are then used to compute the photometric-bias and paired-difference losses,
whose gradients propagate through the denoised images and update only LoRA parameters. 

The total fine-tuning objective is

\begin{equation}
    \mathcal{L}_{\mathrm{LoRA-N2V}} = \mathcal{L}_{\mathrm{pb}} + \mathcal{L}_{\mathrm{pd}}.
\end{equation}
where the two loss terms are of comparable magnitude in our experiments, so no additional weighting is applied. Both flux-aware losses are implemented as mean absolute deviations, making them less sensitive to outliers than squared losses. 
\begin{figure*}
    \centering
    \includegraphics[width=\linewidth]{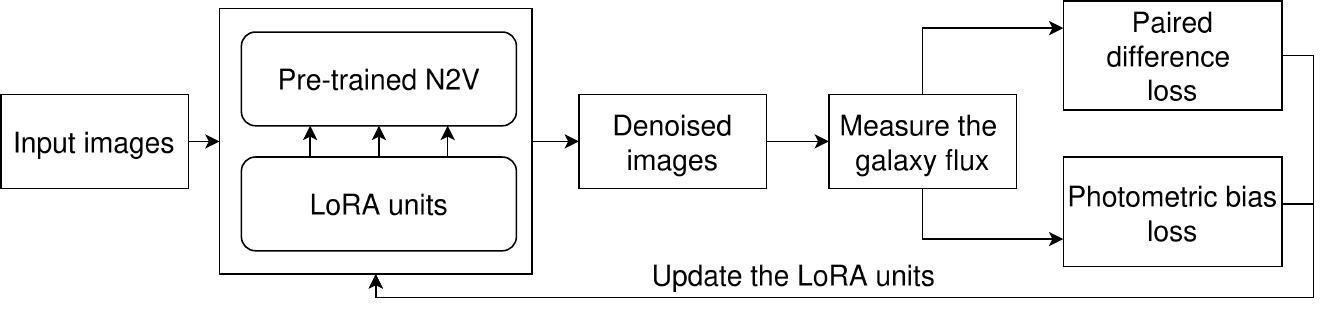}
    \caption{Training procedure of the LoRA-N2V model. A frozen constrained N2V backbone is combined with trainable LoRA modules, and only the LoRA parameters
    are updated during fine-tuning. Denoised images are passed through the differentiable aperture-flux measurement, and the resulting fluxes are used to compute the paired-difference and photometric-bias losses. }
    \label{fig:loratrain}
\end{figure*}

Figure~\ref{fig:loratrain} summarise the LoRA-N2V fine-tuning procedure. The constrained N2V backbone is first trained for 200 epochs on 2000 image stamps, requiring approximately two hours. The LoRA-N2V fine-tuning stage then uses 30\,000 individual image stamps from the NB645 band, corresponding to approximately 15\,000 repeated-exposure pairs randomly selected from
the full set of 295\,837 NB645 measurements. To account for the different photometric regimes, we split the 
sample into bright galaxies with $i_{\mathrm{AB}} < 21$ and faint galaxies with $i_{\mathrm{AB}} \geq 21$, and train separate LoRA-N2V models for the two subsets. 

To accelerate training, we precompute the quantities entering the flux calculation, including the noisy-image fluxes, the background-annulus arrays, and the galaxy-aperture arrays. Since LoRA updates only the low-rank adaptation parameters while keeping the constrained N2V backbone fixed, 200 fine-tuning for epochs takes approximately three hours using a single NVIDIA V100 GPU. Full retraining with the same flux-based losses is estimated to require more than 50 hours.

\section{Denoising results and evaluation}\label{sec:4}

This section evaluates the standard N2V, constrained N2V and LoRA-N2V models. We first test the standard N2V on simulations, where noiseless reference images allow
direct assessment of image quality and flux recovery.
We then apply the three different models to PAUS observations, focusing on flux preservation and repeated-exposure flux consistency. 

\subsection{Image quality and flux bias in simulated images}\label{sec:4.1}

\begin{figure*}[h!]
  \centering
  \includegraphics[width=\textwidth]{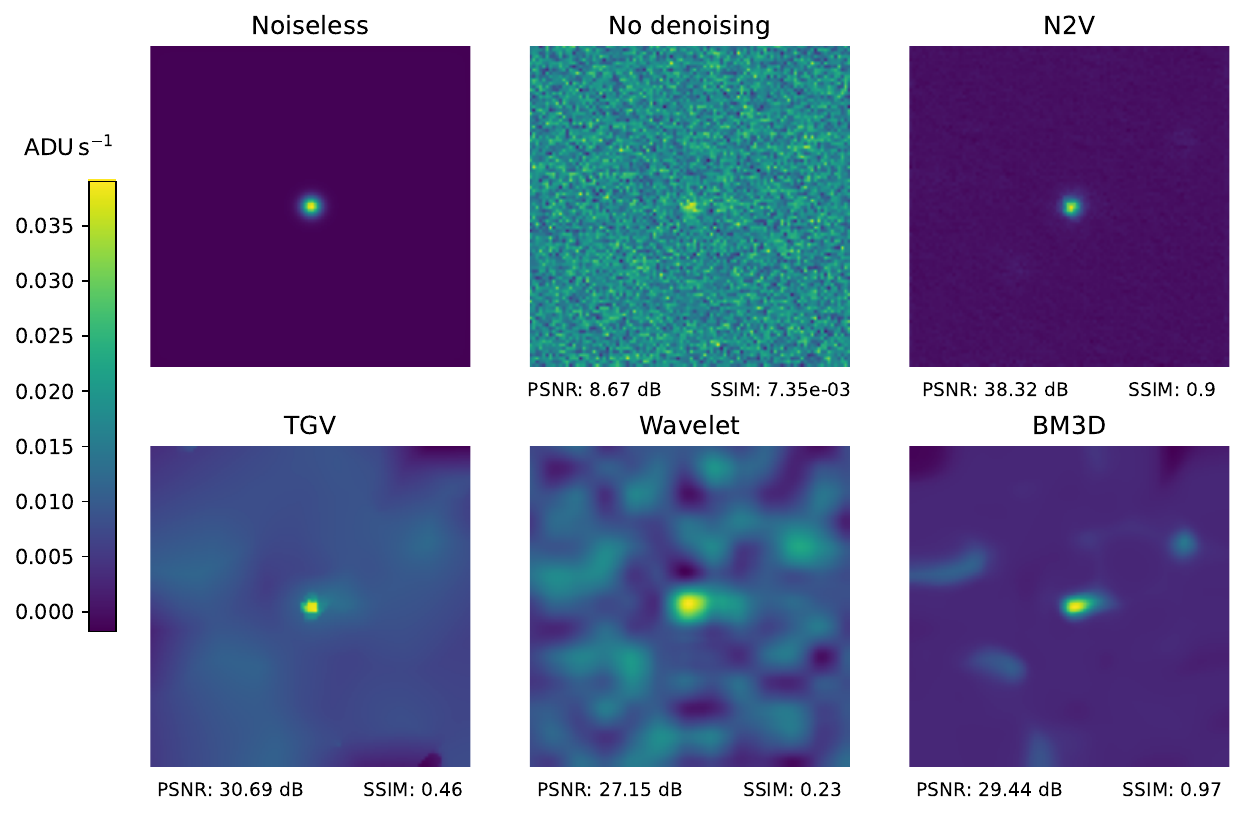}
  \caption{Comparison of N2V and classical denoising baselines on a simulated galaxy stamp. The top row shows, from left to right, the noiseless ground-truth image, the noisy input image, and the N2V-denoised output. The bottom row shows, from left to right, the outputs from TGV, WaveletDictDenoiser, and BM3D. PSNR and SSIM are shown for each denoised image. All images are displayed using the same color scale.}
  \label{fig:Baselines}
\end{figure*}

Figure~\ref{fig:Baselines} compares N2V with the classical denoising baselines introduced in Sect.~\ref{sec:3} for a simulated low-SNR galaxy stamp. 
The N2V output suppresses much of the background noise while preserving the main galaxy features, producing a smooth background that most closely resembles the noiseless reference. By contrast, the classical methods leave more visible residual structure: TGV produces a patchy texture, WaveletDictDenoiser introduces large-scale blotches, and BM3D produces localised smudges.

To quantify denoising performance on simulations, we use three image-quality diagnostics: peak signal-to-noise ratio (PSNR), structural similarity index measure (SSIM), and the standard deviation (STD) of the background pixels. Since noiseless reference images are available, PSNR and SSIM are computed between the denoised and reference images using \texttt{skimage.metrics} \citep{scikit-image}.

\begin{table}[h!]
  \centering
  \caption{Average denoising metrics on the simulated test set. PSNR and SSIM are computed with respect to the noiseless reference images, while STD denotes the standard deviation of background pixels. Arrows indicate whether higher or lower values are better; the best value in each column is shown in bold.}
  \label{tab:average_results}
  \begin{tabular}{l c c c}
    \toprule
    Method & PSNR $\uparrow$ & SSIM $\uparrow$ & STD $\downarrow$ \\
    \midrule
    No denoising & 8.31 & 0.0064 & 1.45$\times 10^{-3}$ \\
    TGV & 29.94 & 0.44 & 6.55$\times 10^{-4}$ \\
    Wavelet & 27.39 & 0.25 & 1.52$\times 10^{-3}$ \\
    BM3D & 29.36 & \textbf{0.96} & $\mathbf{1.73 \times 10^{-5}}$\\
    \textbf{N2V (This work)} & \textbf{38.13} & 0.93 & 2.90$\times 10^{-4}$ \\
    \bottomrule
  \end{tabular}
\end{table}

Table~\ref{tab:average_results} summarises the average metrics over the test set. N2V achieves the highest PSNR and preserves the main galaxy structure without the pronounced artefacts seen in some of the classical reconstructions. Although BM3D gives the highest SSIM and the lowest background standard deviation, Fig.~\ref{fig:Baselines} shows structured artefacts in the BM3D reconstructed image. This illustrates that global image-quality metrics can favour overly smooth or structured reconstructions that are not necessarily optimal for photometry \citep{breger2025study}. We therefore use these metrics only as image-quality diagnostics and complement them below with direct tests of flux preservation.
 
\begin{figure}%[htbp]
    \centering
        \includegraphics[width=1\linewidth]{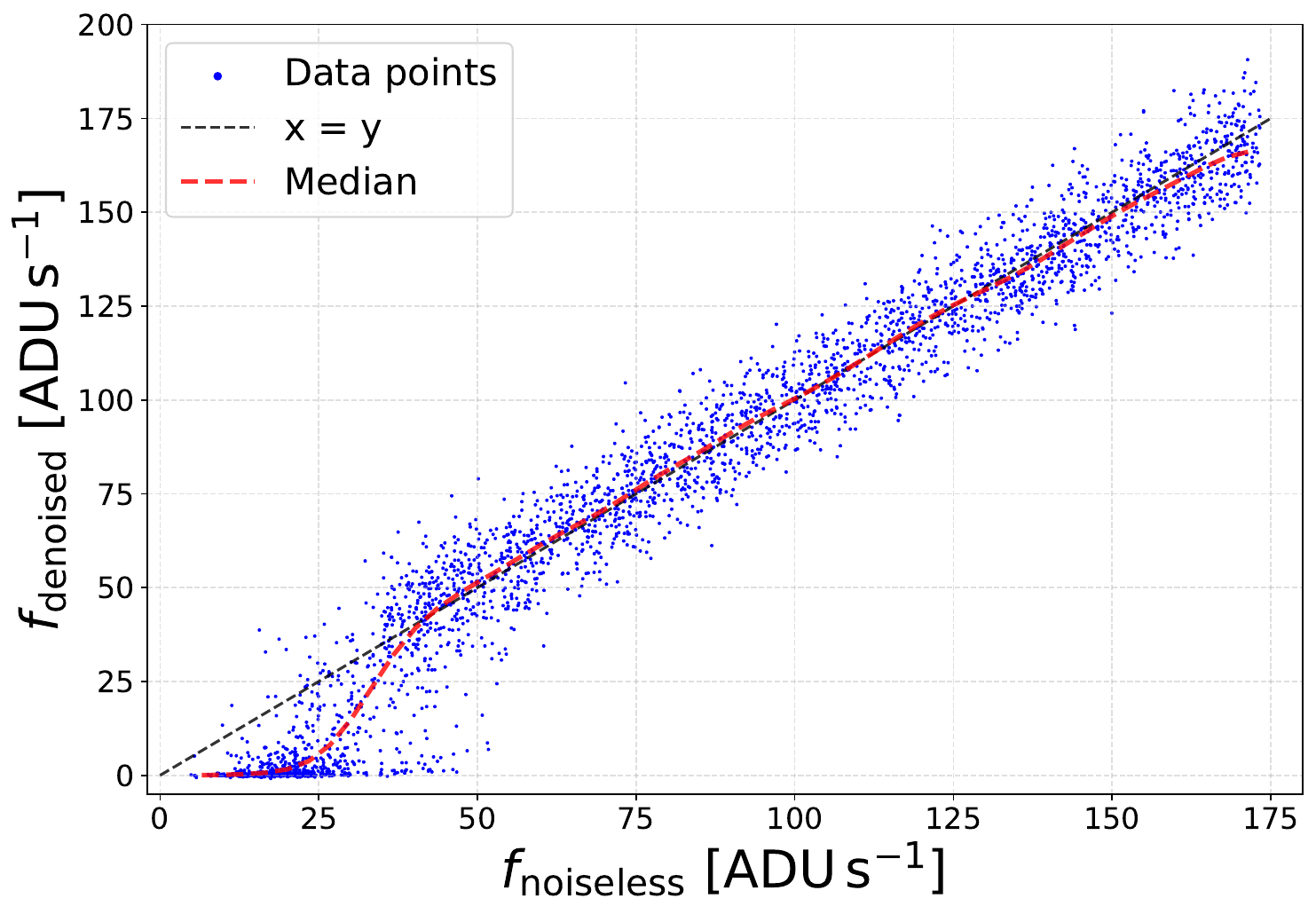}
    \caption{Comparison between the noiseless simulated galaxy flux and the flux recovered after denoising with standard N2V. The horizontal axis shows the ground-truth flux, while the vertical axis shows the flux measured from the denoised stamps.}
    \label{fig:bright}
\end{figure}

We next test whether the improved image quality is accompanied by accurate flux recovery in the simulations. 
Since PAUS catalogue apertures are not available for the simulated images,  we estimate the galaxy flux from the
central 12 × 12 pixel region of each stamp. As shown in Fig.~\ref{fig:bright}, N2V broadly preserves the fluxes of brighter sources, but some low-flux galaxies show substantial flux suppression, in some cases approaching complete loss of the recovered signal. This indicates a
flux-dependent bias, most clearly affecting faint sources.

\begin{figure*}
    \centering
    \includegraphics[width=\linewidth]{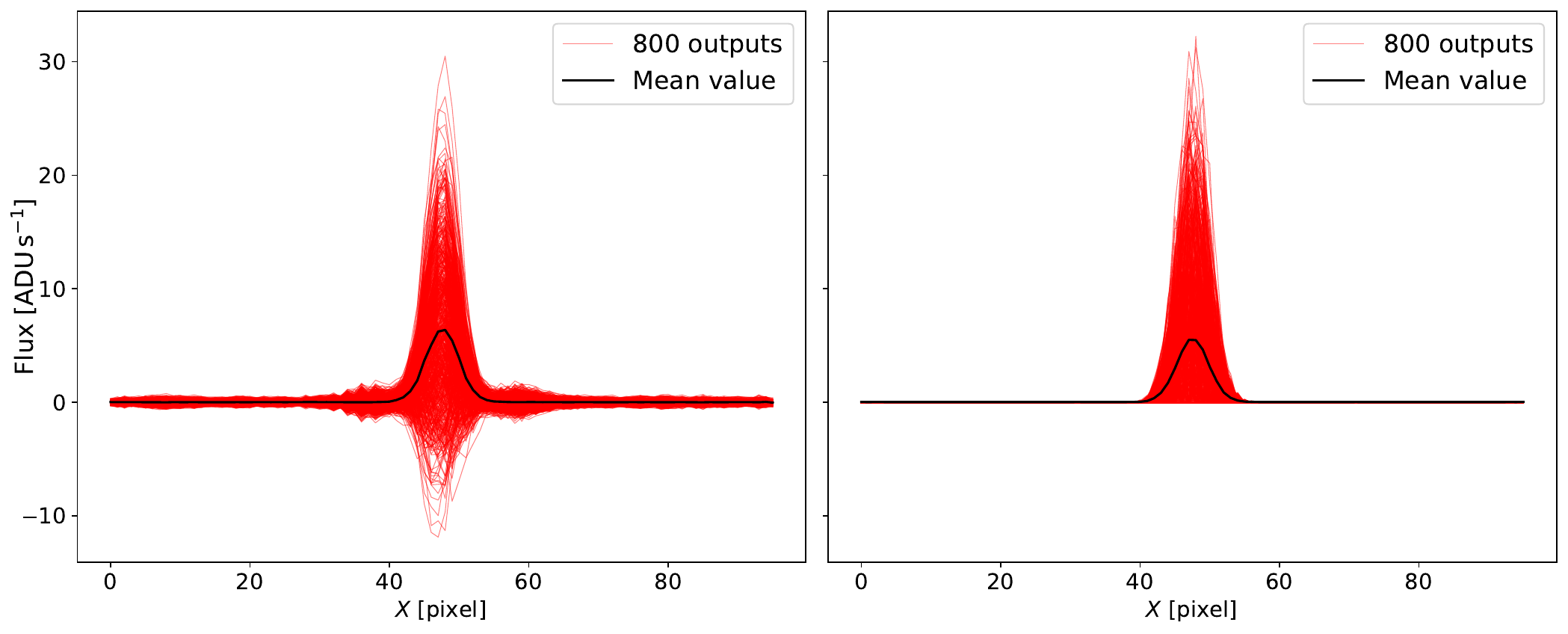}
    \caption{Central-row intensity profiles for 800 stochastic predictions of the same simulated galaxy. \emph{Left:} Standard N2V. \emph{Right:} Constrained N2V. Each curve corresponds to one stochastic prediction. The constrained model suppresses the large negative output values that appear in the standard N2V predictions.}
    \label{fig:800s}
\end{figure*}

To better understand this behaviour, we inspect the central-row profiles of 800 stochastic predictions for
a simulated galaxy, shown in Fig.~\ref{fig:800s}. For standard N2V (left panel), the mean profile remains positive near the source centre, but many individual predictions become negative. Averaging these predictions shifts the recovered profile downward and can therefore
underestimate the flux, especially for weak galaxy signals. This motivated the constrained N2V modification, which is evaluated in the next subsection.

\subsection{Flux preservation with constrained N2V}
The right panel of Fig.~\ref{fig:800s} shows the 800 stochastic predictions obtained with constrained N2V for the same simulated galaxy shown in the left panel for standard N2V.
Compared with standard N2V, the constrained model suppresses large negative outputs and reduces the downward shift in the average profile,  supporting the interpretation that negative stochastic predictions contribute to flux loss observed for faint simulated galaxies.

We next test whether similar flux loss appears in PAUS
observations. Using all galaxies observed in the NB645 band, we compare fluxes before and after denoising with standard and constrained N2V. For each measurement, we define the flux ratio 
\begin{equation}
    R_f \equiv \frac{f_{\mathrm{denoised}}}{f_{\mathrm{original}}}.
\end{equation}

\noindent
where $f_{\mathrm{original}}$ and
$f_{\mathrm{denoised}}$ are measured from the original and denoised PAUS images using the same photometric procedure.
 
\begin{figure}
    \centering
    \includegraphics[width=\linewidth]{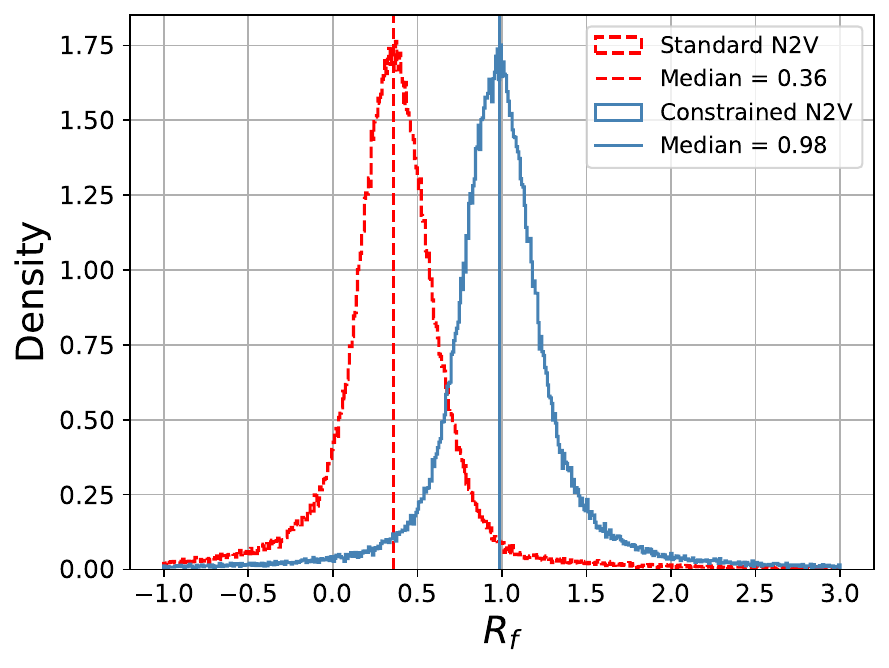}
    \caption{Distribution of flux ratios, $R_f$, for faint PAUS galaxies with $21.5 \leq i_{\mathrm{AB}} < 23$ in the NB645 band.  The red histogram shows standard N2V, while the blue histogram shows constrained N2V. The unbiased expectation is unity.}
    \label{fig:rf_hist}
\end{figure}

The PAUS catalogue spans $17 \leq i_{\mathrm{AB}} \leq 23$. Since the simulations 
show that flux loss is strongest for low-flux sources, we focus on the faintest galaxies, $21.5 \leq i_{\mathrm{AB}} < 23$.  Figure~\ref{fig:rf_hist} shows the corresponding $R_f$ distribution. For standard N2V, the distribution peaks at approximately 0.36, well below the unbiased expectation of unity, indicating a substantial systematic suppression of the fluxes of faint PAUS galaxies. In contrast, constrained N2V peaks at approximately 0.98, showing that the output constraint strongly reduces the systematic flux bias.

\subsection{Photometric performance with LoRA-N2V}

\begin{figure*}
    \centering
    \includegraphics[width=\linewidth]{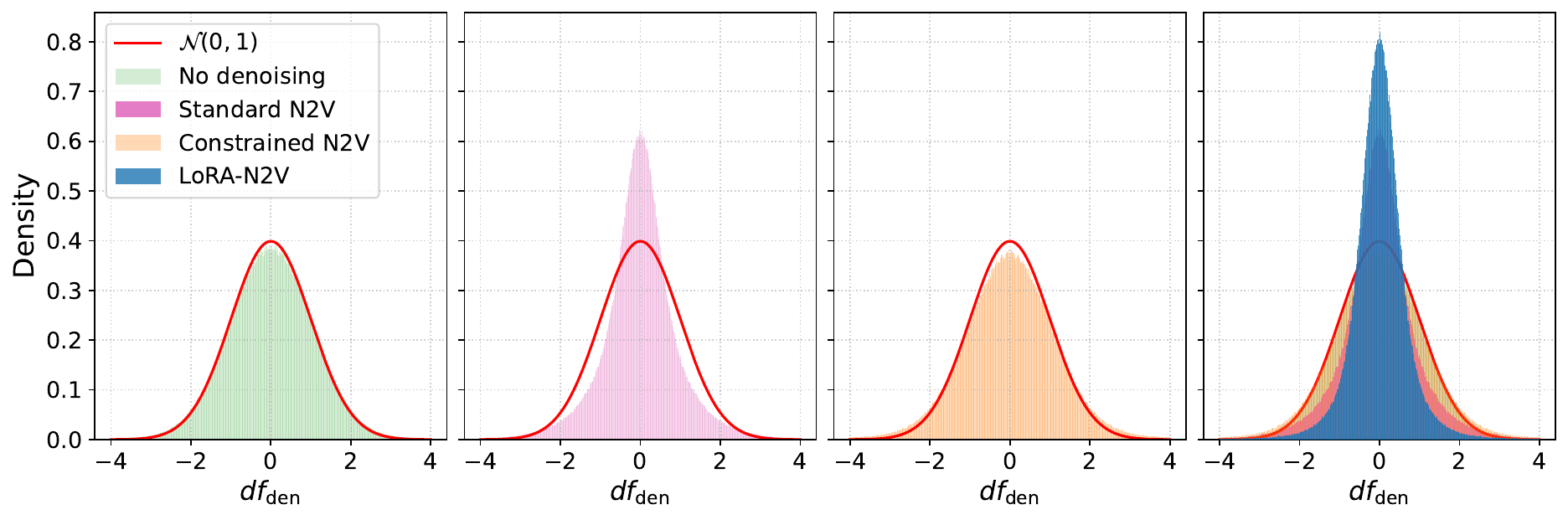}
    \caption{Flux-difference ($df_{\mathrm{den}}$) distributions for PAUS galaxies observed in the NB645 band, using the same
    sample in all cases, $17 \leq i_{\mathrm{AB}} < 23$. From left to right, the panels show the distributions without denoising, standard N2V, constrained N2V, and LoRA-N2V. The final panel overlays the three other distributions for comparison.}
    \label{fig:histdf}
\end{figure*}

We next evaluate whether LoRA-N2V improves the repeatability of PAUS flux measurements. Figure~\ref{fig:histdf} shows the $df_{\mathrm{den}}$ distributions for the NB645 sample; for LoRA-N2V, the histogram is computed only from measurements excluded from training. The no-denoising and constrained N2V cases have similar widths,
indicating that constrained N2V does not substantially reduce the scatter between repeated flux measurements. By contrast, the LoRA-N2V distribution is substantially narrower. 
Since the denominator of $df_{\mathrm{den}}$ is kept fixed to the original PAUS uncertainties, this narrowing indicates reduced repeated-exposure flux scatter relative to the original uncertainty scale. 

To quantify the width of the $df_{\mathrm{den}}$ distributions,  we use $\sigma_{68}$, defined as half the difference between the 84th and 16th percentiles. For a standard normal distribution, $\sigma_{68}$ equals the standard deviation. The no-denoising and constrained N2V cases have $\sigma_{68} \simeq 1$, consistent with the original PAUS uncertainty scale. For LoRA-N2V, $\sigma_{68} = 0.529$, corresponding to an effective reduction in the random flux error of roughly 47\%. For the faintest galaxies, $22 \leq i_{\mathrm{AB}} < 23$, the width decreases further to $\sigma_{68} = 0.415$, indicating a stronger improvement at low-SNR.

\begin{figure}
    \centering
    \includegraphics[width=\linewidth]{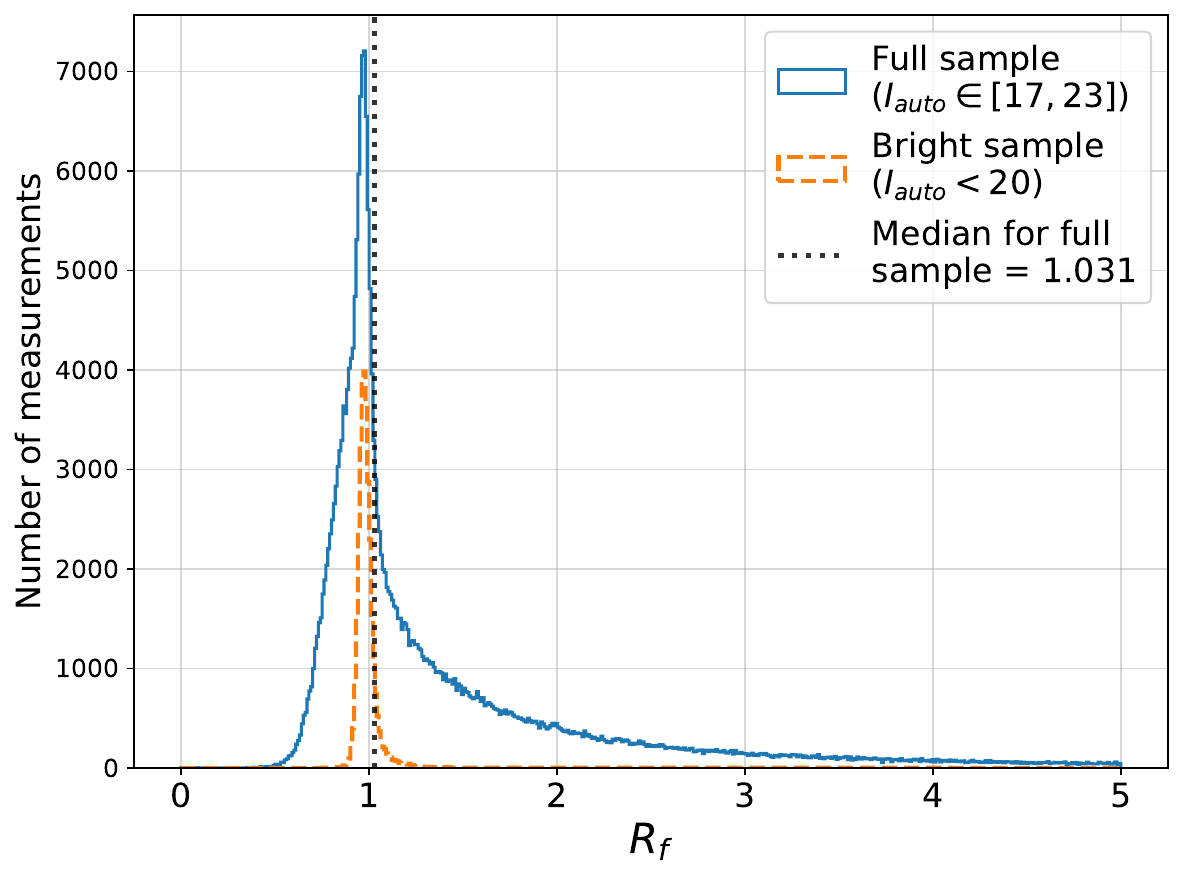}
    \caption{Distribution of flux ratios, $R_f$, after LoRA-N2V denoising in the NB645 band. The blue histogram shows the full sample, while the orange histogram shows the brighter subsample. The vertical dashed line marks the median of the full sample, $R_f=1.031$.}
    \label{fig:loraratio}
\end{figure}

We also test whether LoRA-N2V introduces a systematic flux bias. Figure~\ref{fig:loraratio} shows the distribution of $R_f$ for the full NB645 sample and for a brighter subsample.  The full-sample distribution peaks close to unity but has a pronounced tail towards high $R_f$. This tail is strongly reduced in the bright subsample, whose distribution is narrow and centred near unity,
indicating that the tail is mainly driven by low-flux sources. This is expected: faint galaxies dominate the PAUS sample, with measurements in the range $22 \leq i_{\mathrm{AB}} < 23$ accounting for more than 50\% of the catalogue, and 17.3\% of the NB645 measurements have original fluxes below $1~\mathrm{ADU\,s^{-1}}$. For such sources, small absolute flux changes can produce large values of $R_f$. Because we enforce positive fluxes, the distribution has no left tail.

Because the full-sample $R_f$ distribution has a long tail towards high values, we use the median rather
than the mean as the primary bias diagnostic.
 For the full sample, the median is $R_f = 1.031$, indicating that LoRA-N2V preserves the typical PAUS flux scales to within a few percent, despite the presence of a tail driven mainly by low-flux sources.

Taken together, the $df_{\mathrm{den}}$ and $R_f$ diagnostics indicate that LoRA-N2V substantially improves repeated-exposure flux precision while preserving the median PAUS flux scale to within a few percent. The high-$R_f$ tail, driven mainly
by low-flux sources, indicates that flux ratios
for very faint
galaxies should still be interpreted with care, because small absolute flux changes can produce large ratios when the original flux is close to zero.

\subsection{Multi-band denoising analysis with LoRA-N2V}\label{sec:multi-denoising}

We finally test whether the improvement obtained with LoRA-N2V in the NB645 band extends across the full PAUS narrow-band filter set. For this multi-band analysis, we train and apply LoRA-N2V using images from all 40 filters, restricting the sample to galaxies with $17 \leq i_{\mathrm{AB}} \leq 23$. Because of the size of the full dataset and the computational cost of training including all filters, we use a randomly selected subset comprising 15\% of the available measurements. 
We split this subset into independent training and evaluation samples at the galaxy-pair level, 
ensuring that repeated exposures used for evaluation are not used during LoRA fine-tuning.

\begin{figure*}
    \centering
    \includegraphics[width=\linewidth]{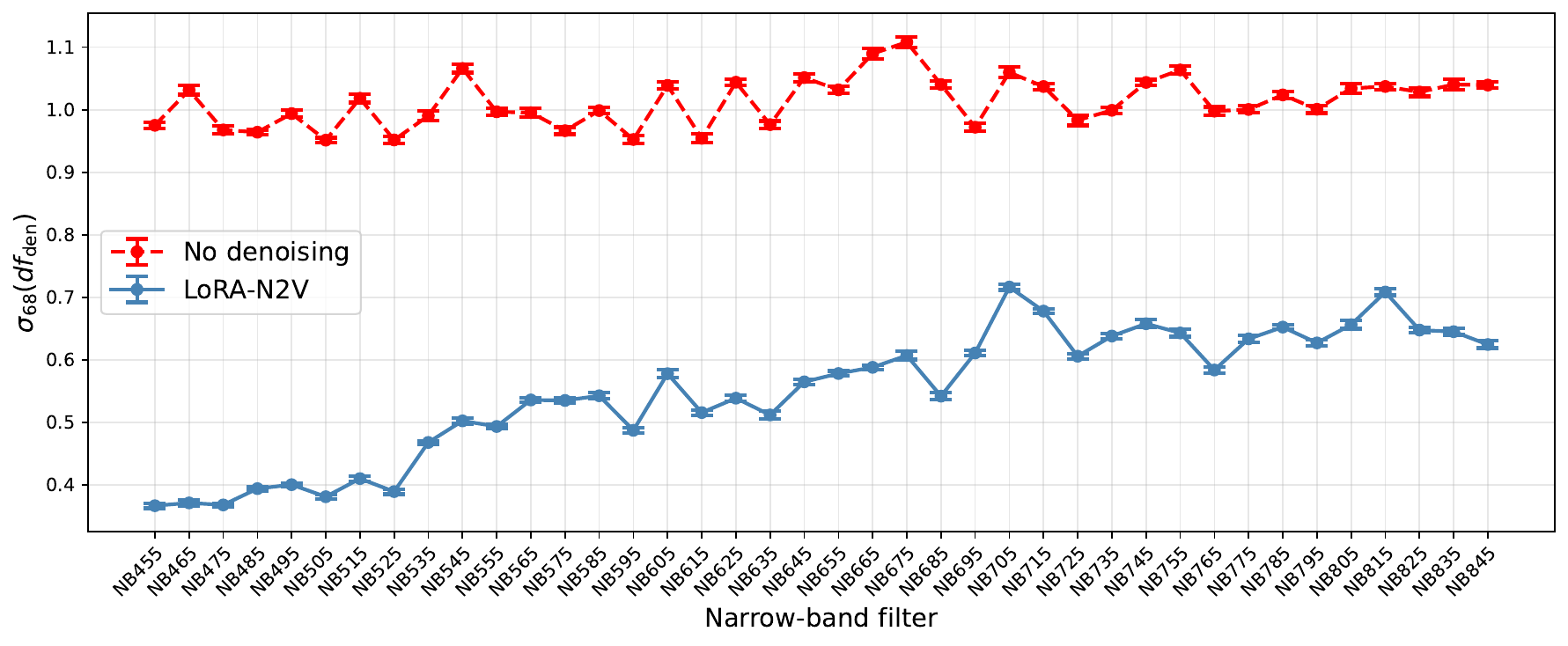}
    \caption{$\sigma_{68}$ of the flux-difference statistic, $df_{\mathrm{den}}$, as a function of the PAUS narrow-band filter. The red curve shows the original measurement without denoising, while the blue curve shows the measurement after LoRA-N2V denoising. Error bars show bootstrap uncertainties. Lower values indicate improved repeated-exposure flux precision relative to the original PAUS uncertainty scale.}
    \label{fig:bandssigma68}
\end{figure*}

After denoising, we compute $df_{\mathrm{den}}$ separately for each filter and summarise the width of each distribution using $\sigma_{68}$. Figure \ref{fig:bandssigma68} compares the resulting values before and after denoising. Across the full wavelength range, LoRA-N2V consistently reduces $\sigma_{68}$ relative to the original measurements. The bootstrap uncertainties are small compared with the observed reduction, supporting that the narrowing
is statistical robust. This shows that the improvement in repeated-exposure flux precision is not restricted to the NB645, but extends across the PAUS narrow-band filter set. 

\begin{figure*}
    \centering
    \includegraphics[width=\linewidth]{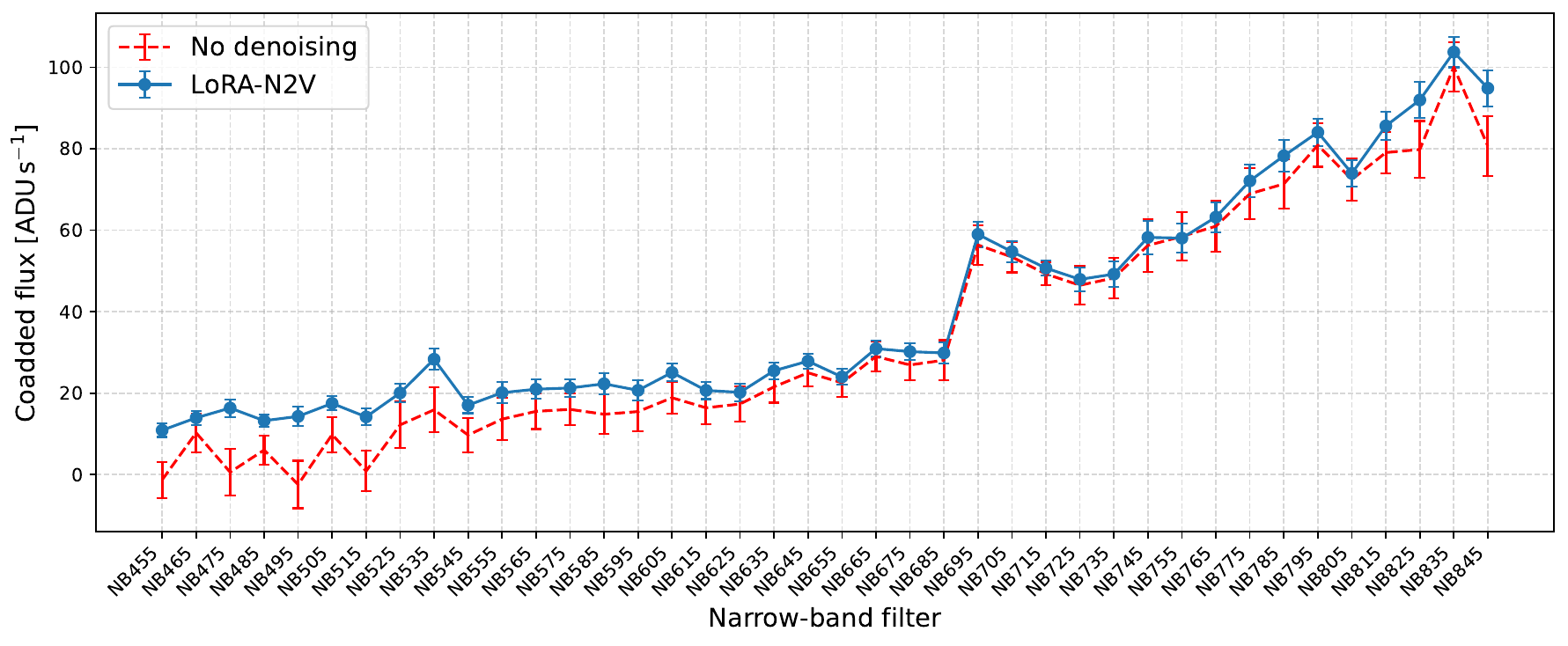}
    \caption{Example coadded narrow-band SED for a PAUS galaxy at $z_{\mathrm{spec}}=0.7264$, before and after LoRA-N2V denoising across the 40 PAUS filters. The redshifted 4000\,\AA{} break is expected between the NB695 and NB705 filters.}
    \label{fig:sed}
\end{figure*}

As an illustrative example, Fig.~\ref{fig:sed} shows the coadded narrow-band SED of a galaxy with spectroscopic redshift $z_{\mathrm{spec}} = 0.7264$, before and after LoRA-N2V denoising. We apply LoRA-N2V to the individual exposures,
measure aperture fluxes, and then coadd the 
denoised fluxes in each band using the same procedure as for the original measurements. The
uncertainties of the denoised coadded SED are
estimated by 
rescaling the original PAUS flux uncertainties in each band by the corresponding $\sigma_{68}$ value measured from the $df_{\mathrm{den}}$ distribution in Fig.~\ref{fig:bandssigma68}. This treats the narrowing of the $df_{\mathrm{den}}$ distribution as an empirical estimate of the reduction in the random flux error relative to the original PAUS uncertainty scale. The original coadded SED shows strong band-to-band fluctuations, especially at shorter wavelengths, whereas the denoised SED is smoother while retaining the main spectral features. 

Galaxies with evolved stellar populations often show a prominent rest-frame spectral break around 4000\,\AA, caused by metal absorption in stellar atmospheres \citep{1983ApJ...273..105B}. For $z_{\mathrm{spec}} = 0.7264$, this feature is 
expected at

\begin{equation}
    \lambda_{\mathrm{obs}} = 4000\,\text{\AA}\, (1 + z_{\mathrm{spec}}) \simeq  6905.6\,\text{\AA}.
\end{equation}
This places the redshifted 4000\,\AA{} break between the NB695 and NB705 filters, consistent with the feature visible in Fig.~\ref{fig:sed}. This example illustrates that LoRA-N2V can reduce narrow-band fluctuations while preserving astrophysically relevant spectral structure.

Further qualitative examples of LoRA-N2V denoising on PAUS image stamps, including repeated exposures and faint catalogue sources, are shown in Appendix~\ref{app:examples}.

\subsection{Downstream application: the D4000 spectral break index}

As an illustrative downstream test, we evaluate the impact of LoRA-N2V on the D4000 spectral break index,  following the methodology of \citet{2022MNRAS.515..146R}. This is intended as a consistency check on the denoised fluxes rather than a dedicated scientific analysis. The D4000 is defined as the flux density ratio between the redward and blueward continuum levels of $4000\,\text{\AA}$ in the rest frame:
\begin{equation}\label{eq:4000}
    \mathrm{D}4000 = \frac{\langle F_{\text{red}} \rangle}{\langle F_{\text{blue}} \rangle},
\end{equation}
where $\langle F_{\text{blue}} \rangle$ and $\langle F_{\text{red}} \rangle$ represent the integrated average flux densities in the continuum regions immediately below and above the $4000\,\text{\AA}$ break, respectively. 

\citet{2022MNRAS.515..146R} computed D4000 directly from the 40 PAUS narrow-band fluxes using a weighted average scheme based on filter overlap and inverse-variance weighting. They found that bright galaxies ($17.0 < I_{\text{auto}} < 21.0$) exhibit a sharp, physical distribution, whereas fainter samples ($17.0 < I_{\text{auto}} < 23.0$) show a noise-driven dispersion that flattens the density distribution and produce unphysical tails. They applied two D4000 definitions: the wide version \citep{1983ApJ...273..105B} ($200\,\text{\AA}$ windows; higher SNR, more dust-sensitive) and the narrow version \citep{1999ApJ...527...54B} ($100\,\text{\AA}$ windows; less dust-contaminated, cleaner age tracer, but more noise-sensitive).

We adopt the same selection criteria as \citet{2022MNRAS.515..146R}. First, we restrict the spectroscopic redshift range to $0.562 < z_{spec} < 0.967$, ensuring that both the blue ($3850\text{--}3950\,\text{\AA}$) and red ($4000\text{--}4100\,\text{\AA}$) rest-frame continuum windows remain fully covered by the PAUS wavelength range. Second, we select galaxies with measurements across all 40 bands in the COSMOS field, giving a sample of 5664 galaxies, of which 755 fall in the bright benchmark sample. These galaxies were excluded from the LoRA fine-tuning sample.

Because $\sigma_{68}$ of the $df_{\mathrm{den}}$ distribution varies across the narrow-band filter set (Fig. \ref{fig:bandssigma68}), the relative inverse-variance weights entering Eq. \ref{eq:4000} depends on which uncertainties are adopted for the denoised fluxes. To avoid introducing a weighting change that could be mistaken for a change in the fluxes themselves, we compute D4000 for both the original and denoised photometry using the original PAUS flux uncertainties, so that the band weights are identical in the two cases, and we use the same weighting method as \citet{2022MNRAS.515..146R}. 

\begin{figure*}
    \centering
    \includegraphics[width=\linewidth]{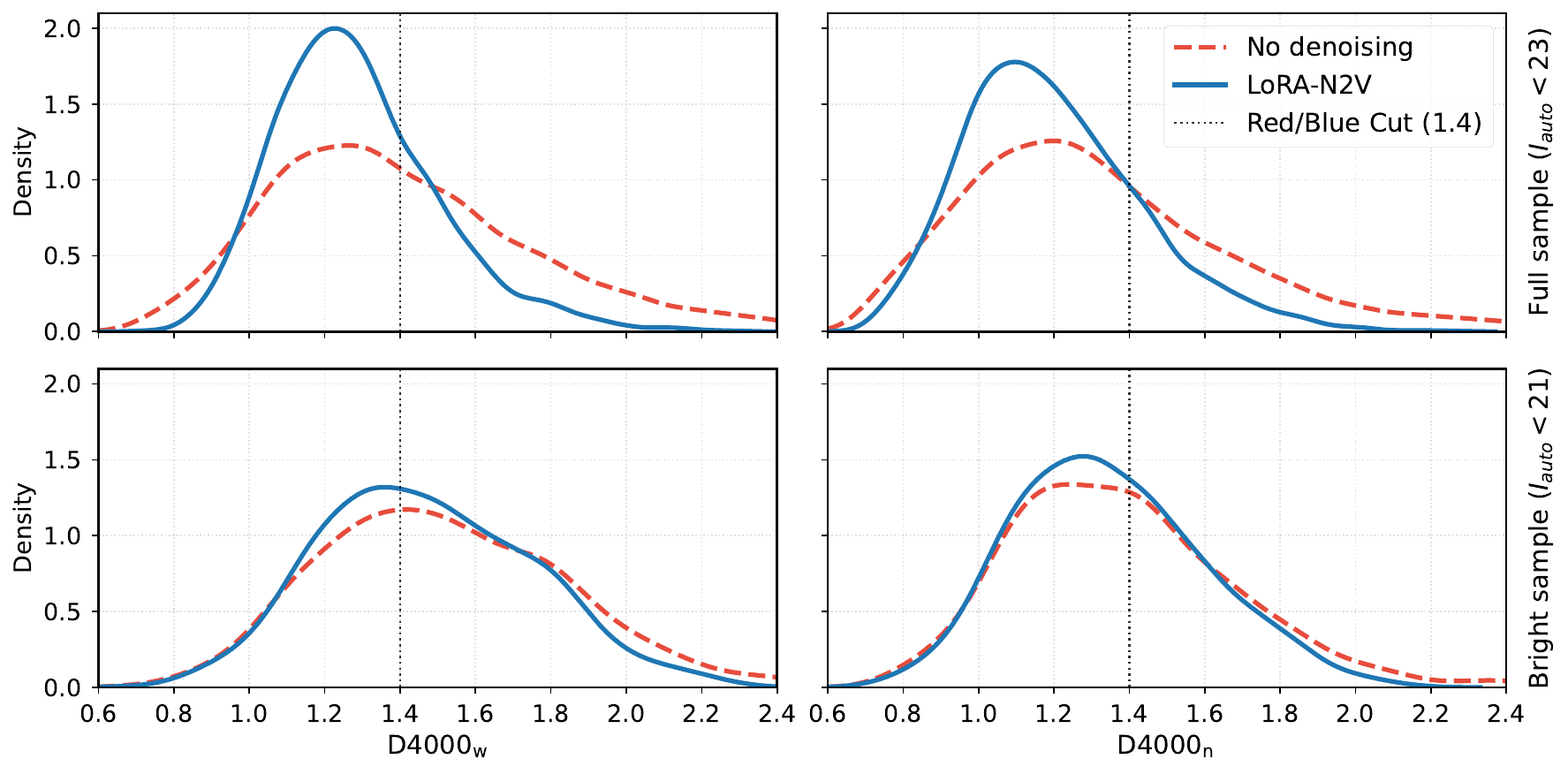}
    \caption{Comparison of rest-frame $\mathrm{D4000_w}$ (left) and $\mathrm{D4000_n}$ (right) density distributions before (orange dashed) and after LoRA-N2V denoising (blue solid). The top panel represents the full sample, while the bottom panel corresponds to the bright benchmark sample. The vertical dotted line indicates the canonical red/blue galaxy threshold at D4000 $= 1.4$.}
    \label{fig:D4000s}
\end{figure*}

Figure~\ref{fig:D4000s} compares the $\mathrm{D4000_w}$ (left) and $\mathrm{D4000_n}$ (right) distributions from no denosing and LoRA-N2V denoised PAUS fluxes, for the full sample (top) and the bright benchmark sample (bottom). For the full sample, the denoised distribution is significantly narrower and the unphysical high-D4000 tail is suppressed. For the bright sample, where the original measurements already have high SNR, the two distributions are nearly identical, with a modest tightening of the peak and no change in the bimodal structure.

Table~\ref{tab:d4000_full_stats} quantifies this contrast. For the bright sample the scatter contracts by 12.5\% ($\mathrm{D4000_w}$) and 14.7\% ($\mathrm{D4000_n}$), with the median shifting downward by 2.1\% and 1.7\% respectively. For the full sample, the improvement is more pronounced: the scatter decreases by 33\% ($\mathrm{D4000_n}$) and 37\% ($\mathrm{D4000_w}$), accompanied by a median shift of $~7\%$ towards lower values.
The reduction in scatter is the expected consequence of the improved repeated flux precision quantified in Sect. \ref{sec:multi-denoising}. The median shift requires more care. Because D4000 is a flux ratio, a multiplicative flux bias common to all bands cancels, but the denoising response is wavelength-dependent (Fig. \ref{fig:bandssigma68}), so a band-dependent residual would not. We therefore measure the median flux ratio $R_f$ separately in each narrow band, finding a monotonic tilt across the spectral range ($R_{f,\,\text{blue}} \approx 0.85$ to $R_{f,\,\text{red}} \approx 0.98$), representing a $14.9\%$ relative difference across the continuum windows). Since the same galaxies enter both measurements, we also compute the per-object ratio of denoised to original D4000, whose median is $0.958$. This isolates the effect of denoising from any change in sample composition. The band-dependence of $R_f$ across the D4000 continuum windows accounts for the observed shift, which should therefore be treated as a residual systematic rather than a physical effect.

Since D4000 is commonly used to separate red and blue populations, we report the practical impact of these changes on classification. Adopting the thresholds of \citet{2022MNRAS.515..146R}, 18.01\% of the full sample and 4.92\% of the bright sample change classification after denoising, with the net flow from red to blue for the wide definition. Similarly, for the narrow definition, 15.40\% of the full sample and 5.43\% of the bright sample change from red to blue. Users of denoised PAUS fluxes for population-level classification should account for this when adopting a fixed threshold.

\citet{2022MNRAS.515..146R} showed that the degradation of the PAUS photometric D4000 distribution relative to spectroscopy is fully accounted for by Gaussian photometric noise: adding scatter of the appropriate magnitude to the VIPERS \citep{2013A&A...557A..54D} D4000 values reproduces the observed PAUS distribution closely in both their samples. The direction of our result is the reverse, since the denoised distribution is narrower and the high-D4000 tail is reduced, which is what would be expected if denoising lowers the effective photometric error entering the D4000 estimator. Establishing that the denoised distribution is genuinely consistent with the same noise model at a reduced error level, rather than merely narrower, would require comparing it against the forward prediction at the measured $\sigma_{68}$ values. Such a comparison with VIPERS, however, is not directly possible within the present analysis, since we restrict our study to the COSMOS field, whereas VIPERS covers a distinct, non-overlapping field. Extending the analysis to the PAUS W4 field to enable calibration against VIPERS spectroscopy would require substantial computational re-training, domain adaptation, and systematic fine-tuning to account for field-to-field variations. Consequently, a full cross-field application and validation against VIPERS is beyond the scope of the present study and is left for future work. This would be particularly useful now that VIPERS spectro-photometry is being used for D4000 based analyses \citep{2026arXiv260607298P}.

\section{Discussion and Conclusions}\label{sec:5}

\subsection{Discussion}

Photometric biases introduced by image-restoration techniques have been reported in previous work. For example, \citet{2024A&A...688A...6A} applied a deep-learning-based two-step deconvolution method to recover high-resolution images from the ESO Distant Cluster Survey (EDisCS, \citealt{2005A&A...444..365W}) and found a systematic flux loss in the reconstructed galaxy images. They mitigated this effect using an additional debiasing procedure based on multiresolution support within a sparsity wavelet framework \citep{Starck1995MultiresolutionSA}. In their analysis, the fractional flux loss increased towards fainter magnitudes. This is consistent with the flux-dependent bias that we find for standard N2V, where faint sources are most strongly affected.

We deliberately apply denoising to the reduced PAUS images rather than to raw detector counts. This keeps the denoised products compatible with the existing PAUS photometric pipeline and avoids 
rerunning the full reduction procedure after denoising. 
The drawback is that reduced images have already undergone processing steps such as bias subtraction, flat-field correction, background subtraction, and the removal of cosmic rays and detector defects, which can introduce weak spatial correlations in the noise. As shown in Appendix \ref{app:noisecorr}, such correlations represent a departure from the ideal independent-pixel noise assumption of likelihood-based blind-spot methods such as PN2V, and may contribute to the background artefacts observed in our PN2V tests. This motivated our use of an N2V-based approach rather than the full PN2V likelihood formulation.

Our results show that standard N2V, in its original form, is not sufficient for unbiased faint-source photometry in PAUS. This limitation is not only a consequence of the image-quality metrics commonly used to evaluate denoising methods, but also reflects a deeper mismatch between pixel-level denoising objectives and integrated photometric measurements. In the low-SNR regime, the network must infer the signal from its spatial context, and the learned image prior can favour smoother or lower-flux reconstructions when the galaxy signal is comparable to the noise. A denoised image can therefore look visually cleaner while still introducing systematic flux biases. We address this limitation with a bounded-output modification that limits large negative predictions, together with flux-aware losses that directly target photometric bias and repeated-exposure scatter.

A limitation of the present analysis is that the LoRA-N2V losses use the original PAUS aperture fluxes as the photometric reference. These fluxes are the relevant operational quantities for PAUS analyses, but they are not noiseless ground truth. The method therefore preserves the PAUS photometric scale and improves repeated-exposure consistency, rather than demonstrating recovery of the true astrophysical flux of each source. We must acknowledge that our method recovers the most probable image under the learned prior, not the true one. 

\begin{table*}[htbp]
\centering
\caption{Comparison of rest-frame D4000 statistical properties (Median and Standard Deviation) before and after LoRA-N2V denoising.}
\label{tab:d4000_full_stats}
\begin{tabular}{lcccccccc}
\toprule
\textbf{Sample} & \textbf{Index} & \multicolumn{3}{c}{\textbf{Median}} & \multicolumn{3}{c}{\textbf{Standard Deviation}} \\
\cmidrule(lr){3-5} \cmidrule(lr){6-8}
 & & \text{No denoising} & \text{LoRA-N2V} & \text{Shift (\%)} & \text{No denoising} & \text{LoRA-N2V} & \text{Reduction (\%)} \\
\midrule
Full Sample      & $\mathrm{D4000_w}$ & 1.3459 & 1.2595 & $-6.42\%$ & 0.3458 & 0.2183 & $-36.87\%$ \\
 & $\mathrm{D4000_n}$ & 1.2574 & 1.1676 & $-7.14\%$ & 0.3508 & 0.2358 & $-32.78\%$ \\
\addlinespace
Bright Sample    & $\mathrm{D4000_w}$ & 1.4849 & 1.4535 & $-2.12\%$ & 0.3208 & 0.2806 & $-12.53\%$ \\
 & $\mathrm{D4000_n}$ & 1.3471 & 1.3237 & $-1.74\%$ & 0.2939 & 0.2507 & $-14.72\%$ \\
\bottomrule
\end{tabular}
\end{table*}

\begin{figure*}
    \centering
    \includegraphics[width=\linewidth]{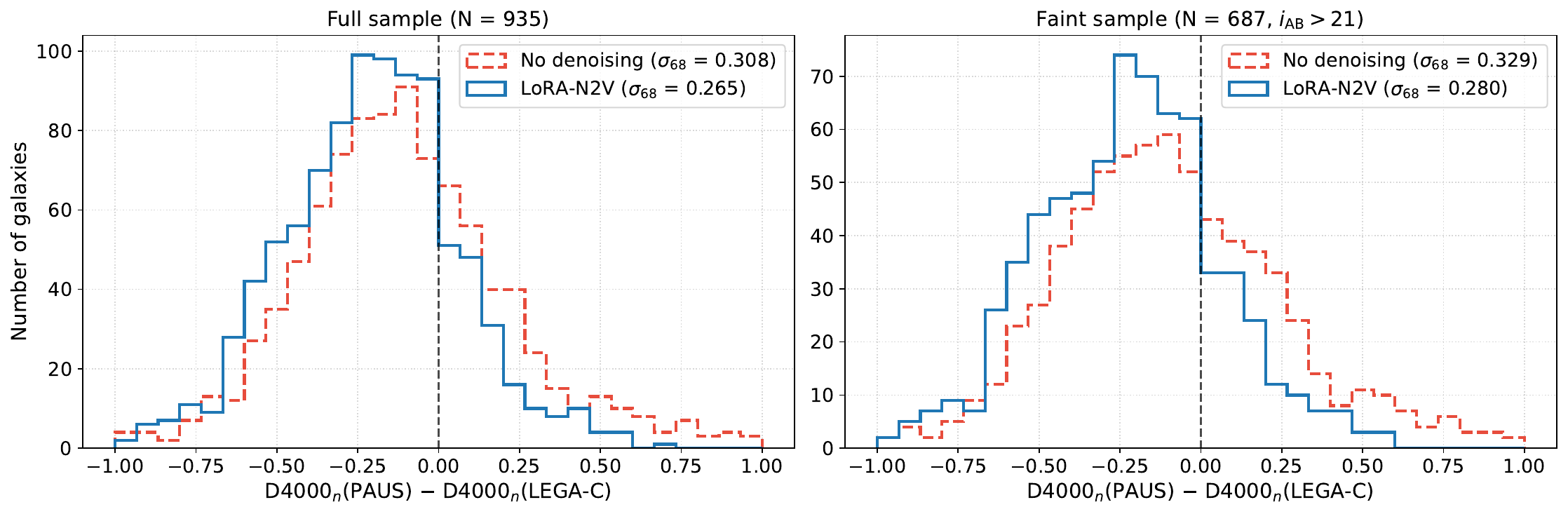}
    \caption{Distribution of $\mathrm{D4000_n}$ residuals, PAUS $-$ LEGA-C, measured before (red) and after (blue) LoRA-N2V denoising. \textit{Left:} full sample ($N = 935$). \textit{Right:} faint subsample ($i_{\mathrm{AB}} > 21$, $N = 687$). Denoising reduces $\sigma_{68}$ from 0.308 to 0.265 (full) and from 0.329 to 0.280 (faint), with no significant change in the median offset.}
    \label{fig:legac}
\end{figure*}

Although the LEGA-C DR3 \citep{2016ApJS..223...29V} spectroscopic footprint directly overlaps with the COSMOS field, a direct absolute comparison of $\mathrm{D4000_n}$ between the two datasets must be interpreted with caution. LEGA-C uses $1.0''$-wide slit spectroscopy, which primarily samples the central regions of galaxies and therefore places greater weight on their older, redder stellar populations. In contrast, PAUS uses total-aperture narrow-band photometry, which also includes contributions from the outer disk and hence from potentially younger and bluer stellar populations. Radial stellar-population gradients can therefore introduce a systematic offset between the $\mathrm{D4000_n}$ measurements. In our comparison, denoising reduces the spread of the PAUS $\mathrm{D4000_n}$ measurements relative to LEGA-C, while a residual systematic offset remains. This combination of reduced scatter and non-zero bias is consistent with improved measurement precision, but prevents the LEGA-C comparison from being interpreted as a direct validation of the absolute $\mathrm{D4000_n}$ calibration. We therefore use it primarily to assess the reduction in scatter produced by the denoising model, which is shown in Fig.\ref{fig:legac}.  

The simulations provide controlled tests with known noiseless references, but they do not capture the full complexity of PAUS background. Constructing
realistic PAUS-like backgrounds with corresponding
noiseless references is difficult:
background cutouts from real PAUS images already contain observational noise, 
while denoising such cutouts to approximate noiseless backgrounds can leave method-dependent residual patterns.
For this reason, PSNR and SSIM are used only on simulations with flat backgrounds, while the main
validation relies on real PAUS observations through
flux ratios, repeated-exposure flux-difference statistics ($df$),
and multi-band consistency tests.

PAUS provides a practical testbed for flux-preserving denoising because it combines low-SNR narrow-band imaging, repeated observations of the same sources, and a well-characterised photometric pipeline. These properties allow denoising methods to be evaluated not only with image-quality metrics, but also with survey-specific flux diagnostics based on repeated exposures. Although applying such methods directly to larger surveys such as \textit{Euclid} would have broader immediate impact, PAUS offers a controlled setting in which the methodology can be developed and tested before being adapted to other datasets.

\subsection{Conclusions}
We tested self-supervised denoising on PAUS narrow-band images, focusing on whether noise can be reduced without biasing photometric measurements. Standard N2V, applied first to simulated galaxy images, improves image-quality diagnostics but introduces a strong flux-dependent bias, with systematic flux underestimation for faint sources.

To mitigate this bias, we introduced a constrained N2V variant with a bounded output that limits large negative predictions in low-SNR regions.
 This substantially reduces the faint-source flux bias. For PAUS galaxies with $21.5 \leq i_{\mathrm{AB}} < 23$ observed in the NB645 band, the standard N2V flux-ratio distribution peaks at $R_f \simeq 0.36$, whereas the constrained N2V distribution peaks close to unity, at $R_f \simeq  0.98$. This shows that the constrained model mitigates the main systematic flux loss introduced by standard N2V.

We introduced LoRA-N2V as a flux-aware fine-tuning strategy for improving repeated-exposure flux precision. The model uses two photometric losses, a paired-difference loss and a photometric-bias loss, while updating only a small number of trainable adaptation parameters and keeping the constrained N2V backbone fixed.
Using the $df_{\mathrm{den}}$ statistic, we find that LoRA-N2V narrows the repeated-exposure flux-difference distribution, with $\sigma_{68}=0.529$ in NB645 and $\sigma_{68}=0.415$ for the faintest galaxies. At the same time, the flux-ratio distribution has a median value of $R_f = 1.031$, indicating the typical PAUS flux scale is preserved to within a few percent.

LoRA-N2V was also tested across the full PAUS narrow-band filter set. The denoised fluxes show consistently lower $\sigma_{68}$ values than the original measurements, indicating improved repeated-exposure flux precision across the wavelength range. An illustrative coadded SED example further shows that LoRA-N2V reduces band-to-band fluctuations while preserving physically interpretable spectral features, such as the redshifted 4000\,\AA{} break. As a downstream application, we estimate the D4000 spectral break index from the denoised fluxes following the method of \citet{2022MNRAS.515..146R}. The measurement scatter decreases by 33–37\% in the faint-dominated full sample, consistent with the improved flux precision.

Overall, our results show that self-supervised denoising can be adapted to astronomical photometry
when 
it is trained with photometry-aware losses and evaluated with photometric diagnostics.
Image-quality metrics alone are insufficient, because visually improved images can still yield biased fluxes. Flux-aware modifications to N2V, combined with efficient LoRA fine-tuning, provide a practical route toward denoising low-SNR survey images while preserving the photometric qualities needed for quantitative analysis. Future work will assess whether this strategy can be transferred to larger photometric surveys with different observing conditions, image characteristics, and photometric pipelines. 

\begin{acknowledgements}
We thank Shaolin Su, Kai Wang, Guangzai Ye, Han-yue Guo, Laura Cabayol Garcia, and Zijian Wang for useful discussions and their kind help. We thank Gabriele Marchello for bringing Noise2Void to our attention in the context of liquid-phase electron microscopy, which helped motivate our later exploration of these approaches for astronomical imaging.
J. Chen was supported by the China Scholarship Council (File No.~202306040042). 
ME acknowledges support from the Spanish Plan Nacional project PID2023-152069NA-I00.
J. Carretero acknowledges support from the grant PID2024-159420NB-C44, funded by MICIU/AEI/10.13039/501100011033 and by ERDF/EU.
CP acknowledges support from the Spanish Plan Nacional project PID2022-141079NB-C32.
JGB acknowledges the support of the Spanish Agencia Estatal de Investigaci\'on through the Research Project PID2024-159420NB-C43 [MICINN-FEDER], and the grant "IFT Centro de Excelencia Severo Ochoa CEX2020-001007-S".
EG acknowledges support from PID2024-156844NB-C21 and PID2022-138896NB from MICINN/MICIU/AEI (/10.13039/501100011033), Maria de Maeztu (CEX2020-001058-M) grant, which include ERDF/FEDER funds from the European Union, and the MaX-CSIC Excellence Award MaX4-SOMMA-ICE.
FJC acknowledges support from PID2022-141079NB-C31 and PID2025-173734NB-C31 from MICINN/MICIU/AEI (/10.13039/501100011033), Maria de Maeztu (CEX2020-001058-M) grant, which include ERDF/FEDER funds from the European Union, and the MaX-CSIC Excellence Award MaX4-SOMMA-ICE.
ISN acknowledges support from PID2024-159420NB-C42.
The data production, processing and analysis tools for this paper have been developed, implemented and operated in collaboration with the Port d’Informació Científica (PIC) data center. 
This work has made use of CosmoHub, developed by PIC (maintained by IFAE and CIEMAT) in collaboration with ICE-CSIC. It received funding from the Spanish government (grant EQC2021-007479-P funded by MCIN/AEI/10.13039/501100011033), the EU NextGeneration/PRTR (PRTR-C17.I1), and the Generalitat de Catalunya.
The PAU Survey is partially supported by MINECO under grants CSD2007-00060, AYA2015-71825, ESP2017-89838, PGC2018-094773, PGC2018-102021, PID2019-111317GB, SEV-2016-0588, SEV-2016-0597, MDM-2015-0509 and Juan de la Cierva fellowship and LACEGAL and EWC Marie Sklodowska-Curie grant No 734374 and no.776247 with ERDF funds from the EU Horizon 2020 Programme, some of which include ERDF funds from the European Union. IEEC and IFAE are partially funded by the CERCA and Beatriu de Pinos program of the Generalitat de Catalunya. Funding for PAUS has also been provided by Durham University (via the ERC StG DEGAS-259586), ETH Zurich, Leiden University (via ERC StG ADULT-279396 and Netherlands Organisation for Scientific Research (NWO) Vici grant 639.043.512), University College London and from the European Union's Horizon 2020 research and innovation programme under the grant agreement No 776247 EWC. The PAU data center is hosted by the Port d'Informaci\'o Cient\'ifica (PIC), maintained through a collaboration of CIEMAT and IFAE, with additional support from Universitat Aut\`onoma de Barcelona and ERDF. We acknowledge the PIC services department team for their support and fruitful discussions.

\end{acknowledgements}

\section*{Data availability}
The PAUS catalogue data are publicly available through the PAUS public data release at \url{https://pausurvey.org/public-data-release/}. The PAUS Master Catalog \citep{2024MNRAS.534.1504N}, available through CosmoHub, is the recommended access point for PAUS photometry and photo-z data. Raw and reduced PAUS images, together with catalogue metadata, are also available through the PAUS public data-release infrastructure. The intermediate files, trained model weights, and denoised flux products generated for this work are available on Zenodo at \url{https://doi.org/10.5281/zenodo.21197885}.

\section*{Code availability}
The code developed in this work is publicly available at \url{https://github.com/KitfungChan/DenoisingPAUS}. This repository includes the training and evaluation code for the denoising models, together with notebooks for reproducing the figures. The PAUS-like simulated images were generated with the simulation pipeline described in Appendix~\ref{app:sim}, which is available at \url{https://github.com/marberi/Flagship4ML}. The forced-photometry implementation used in this work is available at \url{https://github.com/PAU-survey/nayo}.

\bibliographystyle{aa} % style aa.bst
\bibliography{apssamp} % your references Yourfile.bib

\begin{appendix}

\twocolumn
\section{Spatial noise correlations and PN2V behaviour in PAUS images}\label{app:noisecorr}

This appendix tests whether the reduced PAUS images satisfy the spatial-noise assumptions underlying blind-spot denoising methods. We quantify spatial correlations in source-free background regions using autocorrelation measurements and discuss the implications for 
PN2V denoising of PAUS images.

\subsection{Spatial noise correlations}

Blind-spot denoising methods such as N2V and PN2V assume that the 
noise in the masked central pixel cannot be predicted from its surroundings. This assumption is satisfied for spatially uncorrelated noise, but can be weakened by weak correlations introduced during image calibration and reduction. Since we apply the methods to reduced PAUS images, we test this assumption in source-free background regions.

We quantify the background-noise correlations using the two-dimensional autocorrelation function (2D ACF;  \citealt{gonzalez2017digital}). The 2D ACF measures the similarity between an image patch and shifted copies
of itself as a function of spatial displacement. For spatially uncorrelated noise, the normalised ACF is expected to be sharply peaked at zero lag and close to zero elsewhere. Spatially correlated noise instead produces extended or structured features around the central peak.

A total of 20\,000 source-free PAUS background patches of size $31\times31$ pixels are randomly selected from regions away from the image borders, avoiding masked areas, detected sources, and obvious artefacts.
For each patch, the mean background level is subtracted and 
the 2D autocorrelation is computed using \texttt{scipy.signal.correlate2d} \citep{virtanen2020scipy}. The
normalised autocorrelation is defined as

\begin{equation}
\rho_{\Delta x,\Delta y}
=
\frac{\mathrm{cov}(X,Y)}{\sigma_X\sigma_Y},
\end{equation}
where $X$ and $Y$ denote the overlapping pixel values in the original and shifted mean-subtracted background patches, respectively. We normalise each correlation map by its zero-lag value and then average the normalised maps over the 20\,000 background patches.

\begin{figure}%[htbp]
    \centering
    \includegraphics[width=1\linewidth]{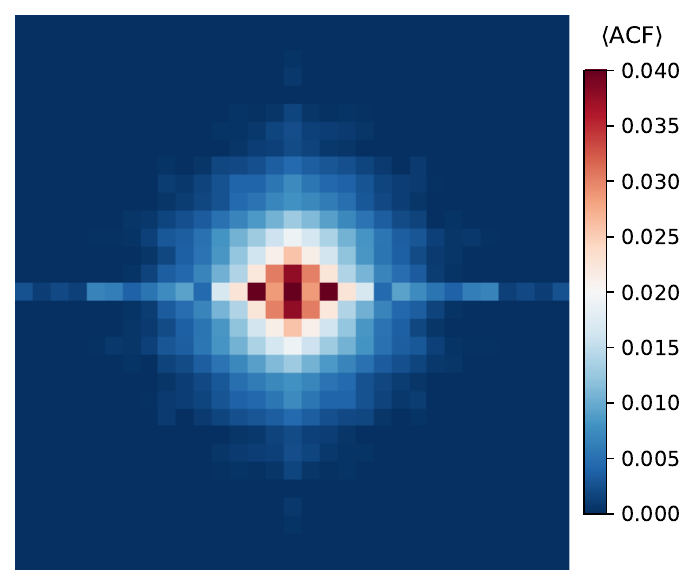}%use pdf
    \caption{Mean normalised autocorrelation map measured from 5000 source-free PAUS background patches. The map is computed on a $31\times31$ lag grid. The colour scale is clipped to $[0, 0.04]$ to highlight weak off-centre correlations; the zero-lag peak is therefore saturated in the display.}
    \label{fig:ACF}
\end{figure}

Figure \ref{fig:ACF} shows the mean normalised correlation map. Around the central peak, the map shows an extended
low-amplitude positive correlation halo,
together with a weak horizontal feature. 
Negative off-centre values are much smaller
in amplitude and are consistent with 
residual statistical noise in the autocorrelation estimate.
This indicates that the noise in the reduced PAUS images is not perfectly spatially uncorrelated. The horizontal structure is consistent with weak correlations introduced by the PAUS reduction pipeline, including the Gaussian filtering applied to correct readout oscillations \citep{2023MNRAS.523.3287S}.

\subsection{Implications for PN2V}
\label{app:noisecorr_impl}

\begin{figure*}%[htbp]
    \centering
    \includegraphics[width=1\linewidth]{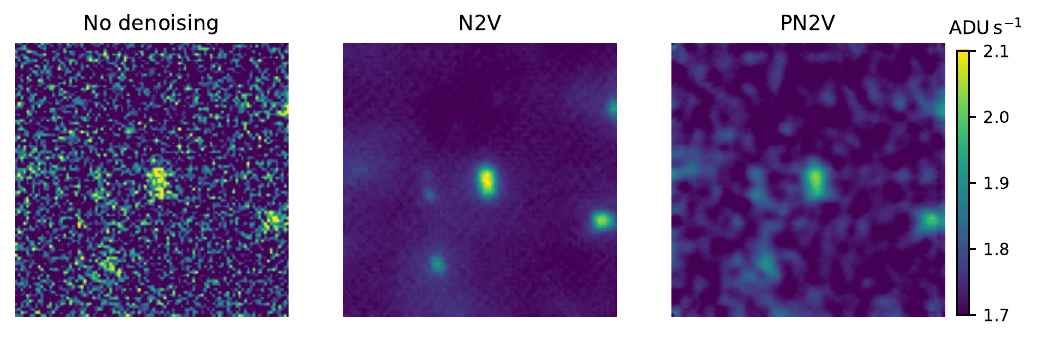}%use pdf
    \caption{Qualitative comparison of denoising results for a PAUS image. \emph{Left:} Original PAUS image. \emph{Middle:} N2V-denoised image. \emph{Right:} PN2V-denoised image using a Poisson noise model. All panels are shown with the same intensity range.}
    \label{fig:N2VPN2V}%use capital of the title.
\end{figure*}

These correlations are small, but they affect blind-spot denoising methods differently. They represent a mild departure from the ideal independent-noise assumption of N2V, while they can be more problematic for PN2V because it also relies on an explicit pixel-level noise likelihood. Even weak deviations from the assumed noise model can therefore lead to structured residuals or background artefacts.

Figure \ref{fig:N2VPN2V} shows a qualitative comparison for a representative PAUS image. The PN2V output, obtained using a Poisson noise model whose parameters were fitted to PAUS data, recovers the central source but leaves visible background artefacts. The N2V output gives a smoother background for the same image and display range. This comparison does not prove that the artefacts are caused by the measured correlations, but it is consistent with the expectation that the full PN2V likelihood is more sensitive to deviations from the assumed noise model.

For this reason, we do not use the full PN2V likelihood formulation in the final analysis. Instead, we adopt a N2V-based approach and address photometric preservation through the constrained output and flux-aware LoRA fine-tuning described in Sect.~\ref{sec:3}.

\section{Pipeline for simulating PAUS images}\label{app:sim}

The PAUS-like simulations provide controlled image data for testing denoising performance and flux preservation. They include the main observational ingredients relevant for these tests: catalogue photometry, galaxy morphology, zero point, exposure time, PSF convolution, background model, and noise realisation. All simulated stamps have  size $96\times96$ pixels, matching the image cutouts used for training and evaluation.

Mock galaxy images are generated by combining photometric information  and morphological parameters through the following steps:

\begin{enumerate}
    \item \textit{Image grid and instrument setup.} 
    Each simulated image is generated on a fixed $96\times96$ pixel grid. To minimise pixelisation effects during profile generation and PSF convolution, the galaxy models are first created on an oversampled grid and then rebinned to the target pixel scale.

    \item \textit{Flux assignment}.  
    The total flux in each band is computed from the input photometry, exposure time, and zero point. The flux is then split between bulge and disk components according to the bulge fraction,

\begin{align}
F_{\mathrm{bulge}} &= F_{\mathrm{tot}}\, f_{\mathrm{bulge}} \\
F_{\mathrm{disk}} &= F_{\mathrm{tot}}\, (1 - f_{\mathrm{bulge}})
\end{align}

where $f_{\mathrm{bulge}}$ denotes the fraction of the total flux assigned to the bulge component.

\item \textit{Galaxy profile generation.} 
The bulge and disk components are represented by elliptical S\'{e}rsic profiles \citep{sersic1968atlas,Graham_Driver_2005}, generated with \texttt{Sersic2D} from \texttt{astropy.modeling} \citep{astropy:2013,astropy:2018,astropy:2022}. The bulge is modelled with a high S\'{e}rsic index, typically close to $n=4$, while the disk is modelled with $n=1$. The half-light radii, ellipticities, position angles, and centroids are taken from the input morphology catalogue. Each component is normalised so that its integrated flux matches the assigned value.

\item \textit{PSF convolution.} 
The simulated galaxy images are convolved with a circular Moffat PSF \citep{1969A&A.....3..455M,1993MNRAS.264..961S} generated with \texttt{astropy.modeling.Moffat2D}. We use $\alpha=1$ and $\beta=4.76$ for the Moffat profile.

\item \textit{Background model.}
A spatially uniform background is added to each stamp, with background levels sampled from a distribution calibrated to PAUS observations.

\item \textit{Noise realisation and image units.}
Poisson noise is applied in count space after PSF convolution and background addition. The noise images are then divided by the exposure time so that the final simulated stamps are expressed in the same units as the reduced PAUS images, $\mathrm{ADU}\,\mathrm{s}^{-1}$.
\end{enumerate}

\subsection{Zero point and exposure time}

The zero point, $ZP$, sets the conversion between calibrated magnitude and measured count rate. We use the convention

\begin{equation}
    m = ZP - 2.5 \log_{10}\!\left(\frac{C}{t_{\mathrm{exp}}}\right),
\end{equation}
where $m$ is the calibrated magnitude, $C$ is the total measured count, and $t_{\mathrm{exp}}$ is the exposure time. The expected count is therefore
\begin{equation}
C = t_{\mathrm{exp}}\,10^{-0.4(m-ZP)} .
\end{equation}

\begin{figure}[h]
    \centering
    \includegraphics[width=1\linewidth]{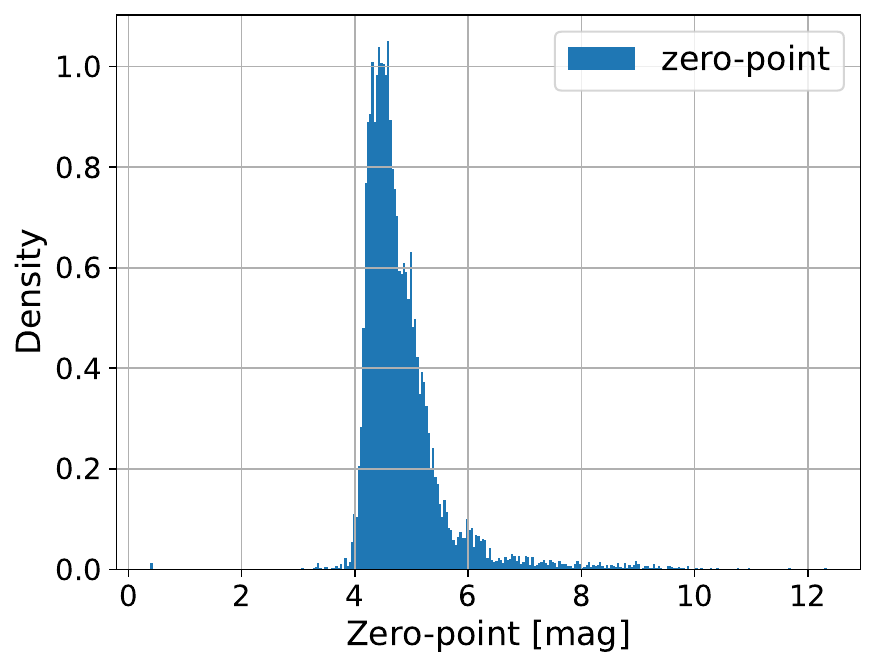}%use pdf
    \caption{Empirical distributions of PAUS zero points, $ZP$, used in the image simulations. The distribution is derived from PAUdm measurements in the COSMOS field across the 40 narrow-band filters and is sampled using kernel density estimation.}
    \label{fig:simulation_zp}
\end{figure}

In the simulations, $ZP$ values are sampled from an empirical PAUS distribution derived from the image-level zero points in the PAUdm for the COSMOS field across the 40 PAUS narrow-band filters  \citep{2007ApJS..172....1S, 2021MNRAS.501.6103A,2024MNRAS.531.5067C}. The distribution, shown in Fig.~\ref{fig:simulation_zp}, is modelled
using kernel density estimation with \texttt{scipy.stats} \citep{virtanen2020scipy}.
The gain is set to $0.71~e^{-}\,\mathrm{ADU}^{-1}$ \citep{2019AJ....157..246P}, consistent with the effectively constant gain in the PAUS data. The exposure time is assigned according to the PAUS filter tray, taking values of 70, 80, 90, 110, 120, or 130 s. 

\subsection{Noise modelling}
Photon-counting noise is applied in count space, before converting to $\mathrm{ADU}\,\mathrm{s}^{-1}$. For each pixel with expected count $\lambda$, the noisy count is drawn as

\begin{equation}
N \sim \mathrm{Poisson}(\lambda).
\end{equation}
The resulting noisy images are then divided by the exposure time to match the units of the reduced PAUS images.

These simplified simulations do not model all detector-level or reduction-induced noise components, such as read noise, and correlated noise from image processing. The impact of these additional effects is instead assessed using real PAUS images.

\section{Interpretation of the flux-difference statistic}\label{app:df}

The flux difference statistic, $df$, defined in Eq.~\ref{eq:df}, compares two
flux measurements of the same galaxy obtained in the same filter
but from different exposures. This appendix summarises its expected behaviour in the ideal Gaussian case and illustrates why it must be interpreted together with a flux-bias diagnostic.

In the ideal case, two flux measurements of the same galaxy can be written as
\begin{equation}
    f_1 = f_{\mathrm{true}} + \epsilon_1 \sigma_1, \qquad
    f_2 = f_{\mathrm{true}} + \epsilon_2 \sigma_2,
\end{equation}
where $f_{\mathrm{true}}$ is the true galaxy flux, $\sigma_1$ and $\sigma_2$
are the corresponding flux uncertainties, and $\epsilon_1$ and $\epsilon_2$ are
independent standard normal variables. Substituting these expressions into Eq.~\ref{eq:df} gives
\begin{equation}
    df_{\mathrm{orig}} = \frac{\epsilon_1 \sigma_1 - \epsilon_2 \sigma_2}{\sqrt{\sigma_1^2 + \sigma_2^2}}.
    \label{eq:df_expanded}
\end{equation}

The numerator of $df_{\mathrm{orig}}$ is normally distributed with zero mean and variance $\sigma_1^2 + \sigma_2^2$. Therefore, if the reported uncertainties correctly describe the random flux errors,

\begin{equation}
    df_{\mathrm{orig}} \sim \mathcal{N}(0,1).
\end{equation}
In this ideal case, the width of the $df_{\mathrm{orig}}$ distribution tests whether the scatter between repeated measurements is consistent with the quoted flux uncertainties.

For denoised measurements, we use the convention defined in Eq.~\ref{eq:dfden}: the denoised fluxes enter the numerator, while the original PAUS uncertainties are kept in the denominator. If denoising preserves the flux scale and reduces the random flux error by a factor $k$, the denoised measurements can be written as
\begin{equation}
\tilde{f}_1 = f_{\mathrm{true}} + k\epsilon_1\sigma_1,
\qquad
\tilde{f}_2 = f_{\mathrm{true}} + k\epsilon_2\sigma_2 .
\end{equation}
Substituting these expressions into Eq.~\ref{eq:dfden} gives
\begin{equation}
df_{\mathrm{den}} \sim \mathcal{N}(0,k^2),
\end{equation}

\noindent
and the width of the $df_{\mathrm{den}}$ distribution therefore measures the reduction in the random flux scatter relative to the original PAUS scale. If the denominator were instead updated to $k\sigma_1$ and $k\sigma_2$, the distribution would again have unit variance. 

However, a narrower $df_{\mathrm{den}}$ distribution does not,
on its own, demonstrate that the flux precision has improved. A systematic rescaling of the measured fluxes can also reduce the numerator. To illustrate this degeneracy, we perform controlled simulations using fiducial galaxy fluxes from the Flagship-PAUS catalogue. This catalogue
is currently internal to the PAUS collaboration and is not publicly available.

%  in CosmoHub\footnote{\url{https://cosmohub.pic.es/}} \citep{Carretero:2017Om, TALLADA2020100391}.

For each repeated measurement $j=1,2$, we generate a mock flux
\begin{equation}
f_{\mathrm{sim}}^{(j)}
=
m\left(f_{\mathrm{true}} + k\,\epsilon_j\sigma_j\right),
\label{eq}
\end{equation}

\noindent
where $f_{\mathrm{true}}$ is the noiseless flux, $\sigma_j$ is the flux uncertainty, $\epsilon_j\sim\mathcal{N}(0,1)$, $k$ controls the amplitude of the random noise term, and $m$ is a multiplicative flux-bias factor. The corresponding flux-difference statistic is computed using the original uncertainties,

\begin{equation}
df_{\mathrm{sim}} =
\frac{f_{\mathrm{sim}}^{(1)}-f_{\mathrm{sim}}^{(2)}}
{\sqrt{\sigma_1^2+\sigma_2^2}} .
\end{equation}

\begin{figure*}[h]
    \centering
    \includegraphics[width=1\textwidth]{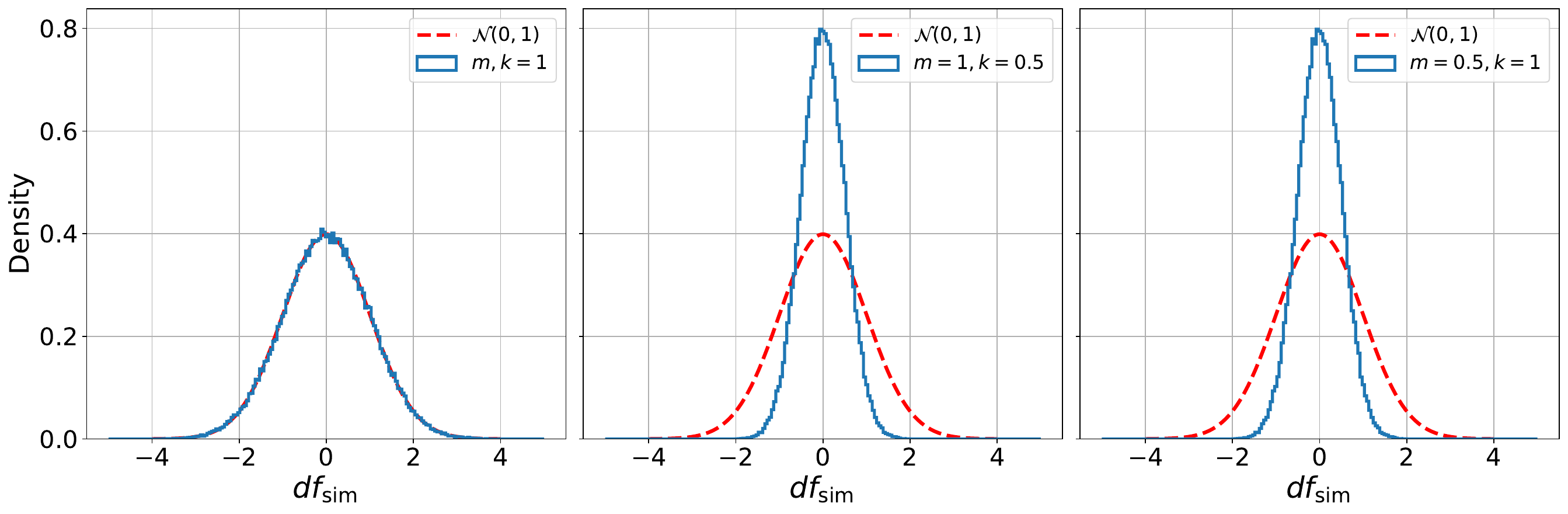}%use pdf
    \caption{Distributions of the flux-difference statistic, $df_{\mathrm{sim}}$, for 
    different combinations of the noise-scaling factor $k$ and the
    multiplicative flux-bias factor $m$.}
    \label{fig:mk}
\end{figure*}

As a baseline, we set $k = 1$ and $m = 1$, corresponding to unbiased flux
measurements with the nominal noise level. The resulting $df_{\mathrm{sim}}$ distribution, shown in the first panel of Fig.~\ref{fig:mk}, closely follows a standard normal distribution.
Setting $m = 0.5$ while keeping
$k = 1$, 
produces a narrower distribution, but this narrowing is caused by the imposed flux suppression rather than by a genuine reduction in random flux error. By contrast, setting $k = 0.5$
and $m = 1$ also narrows the distribution, but preserves the
median flux ratio close to unity, indicating a true reduction
in random scatter.

These tests show that the $df_{\mathrm{den}}$ must be interpreted together with a flux-bias diagnostic. A narrower $df_{\mathrm{den}}$ distribution can indicate improved repeated-exposure precision, but it can also arise from a biased flux rescaling. This motivates the inclusion of a photometric-bias loss in the LoRA-N2V (Sect. \ref{sec:LoRA-N2V}), which penalises changes in the flux scale while allowing the paired-difference loss to reduce repeated-exposure scatter.

\section{PAUS image examples}\label{app:examples}

This appendix provides qualitative examples of denoising applied to PAUS image stamps. These examples illustrate the visual effect of the denoising procedure; the quantitative assessment in the main text is based on aperture-flux measurements, flux-ratio diagnostics, and 
the repeated-exposure $df$ statistics.

\begin{figure*}
  \centering
  \begin{subfigure}[b]{0.33\textwidth}
    \includegraphics[width=\textwidth]{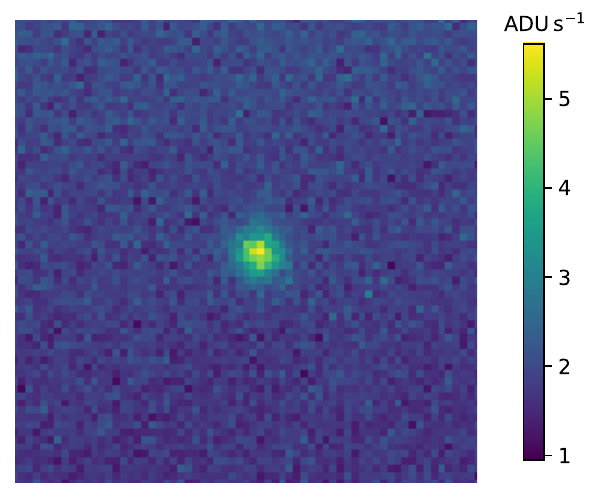}
    \label{fig:denoisecutout1}
  \end{subfigure}
  \hfill
  \begin{subfigure}[b]{0.33\textwidth}
    \includegraphics[width=\textwidth]{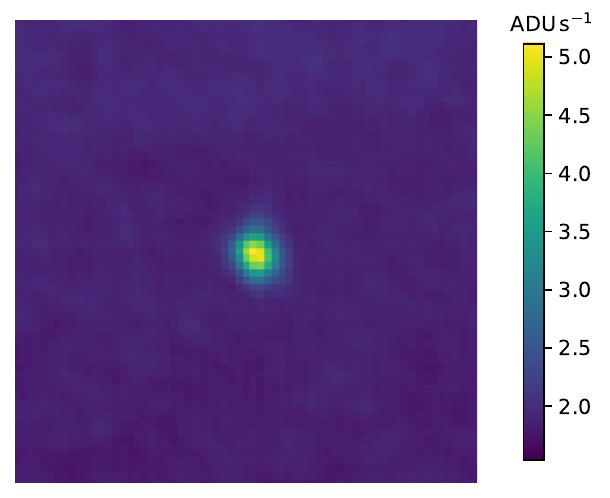}
    \label{fig:denoisecutout2}
  \end{subfigure}
  \hfill
  \begin{subfigure}[b]{0.33\textwidth}
    \includegraphics[width=\textwidth]{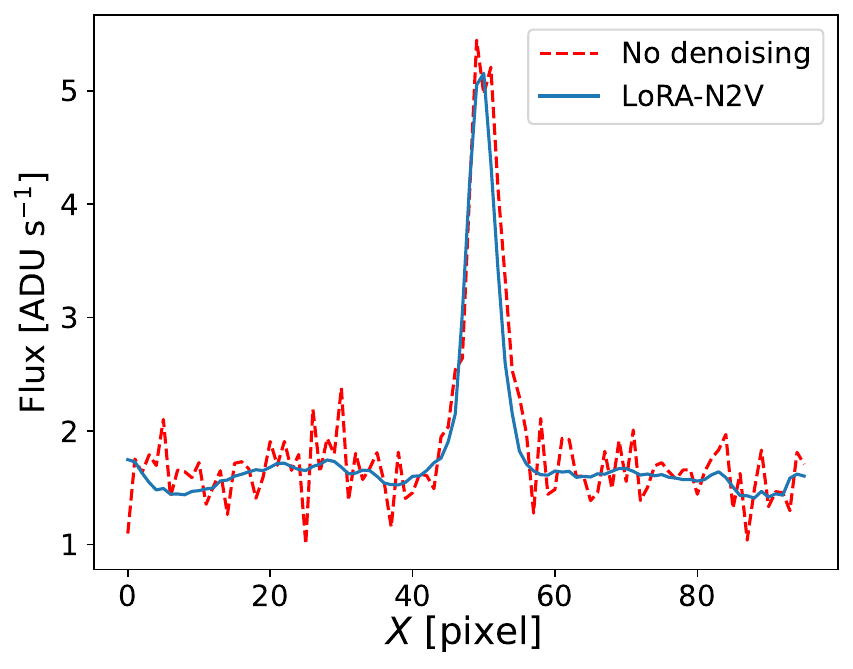}
    \label{fig:profile}
  \end{subfigure}
  \caption{Example of LoRA-N2V denoising applied to a PAUS galaxy cutout. \emph{Left}: Original noisy stamp. \emph{Middle}: Denoised stamp. \emph{Right}: One-dimensional intensity profile across the central galaxy before and after denoising.}
  \label{fig:denoisecutout}
\end{figure*}

Fig.~\ref{fig:denoisecutout} shows an example PAUS galaxy cutout before and after denoising, together with a one-dimensional intensity profile across the central galaxy. Compared to the original noisy image, the denoised image (middle panel) shows reduced noise and a more clearly visible central galaxy. The corresponding profile is smoother after denoising, making the galaxy signal easier to identify visually.

\begin{figure*}
    \centering
    \includegraphics[width=\textwidth]{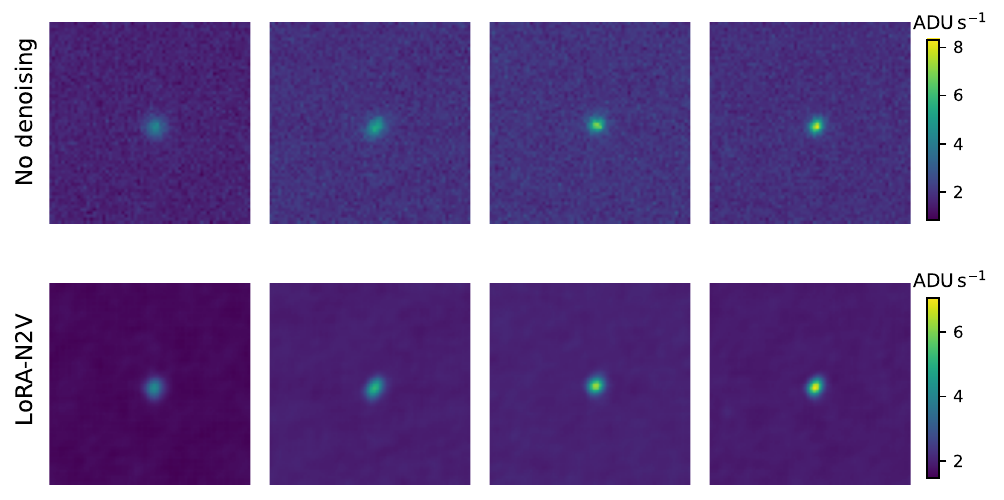}%use pdf
    \caption{Repeated PAUS observations in the same narrow-band filter, with each stamp centred on the same galaxy. The top row shows the original noisy stamp, while the bottom row shows the corresponding LoRA-N2V denoised images. All images are displayed using the same color scale.}
    \label{fig:multiple_exposures_comparison}
\end{figure*}

Figure~\ref{fig:multiple_exposures_comparison} shows repeated observations in the same narrow-band filter. Owing to variations in observing conditions and exposure times, the original stamps exhibit different noise levels and zero points. The corresponding denoised images show reduced small-scale noise and more visually consistent galaxy structures across the repeated observations.

\begin{figure*}
      \centering
      \includegraphics[width=\linewidth]{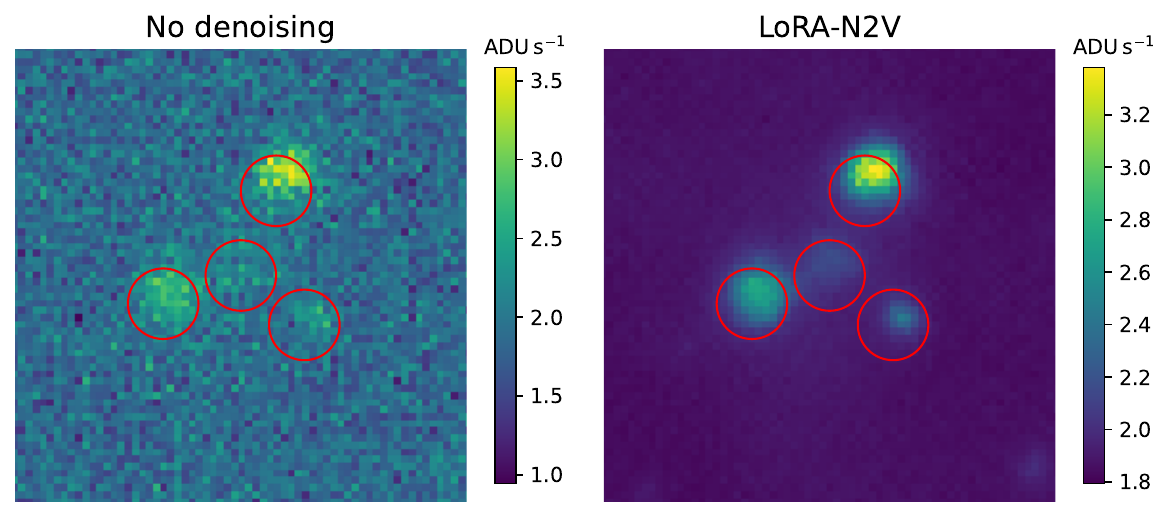}
      \includegraphics[width=1\textwidth]{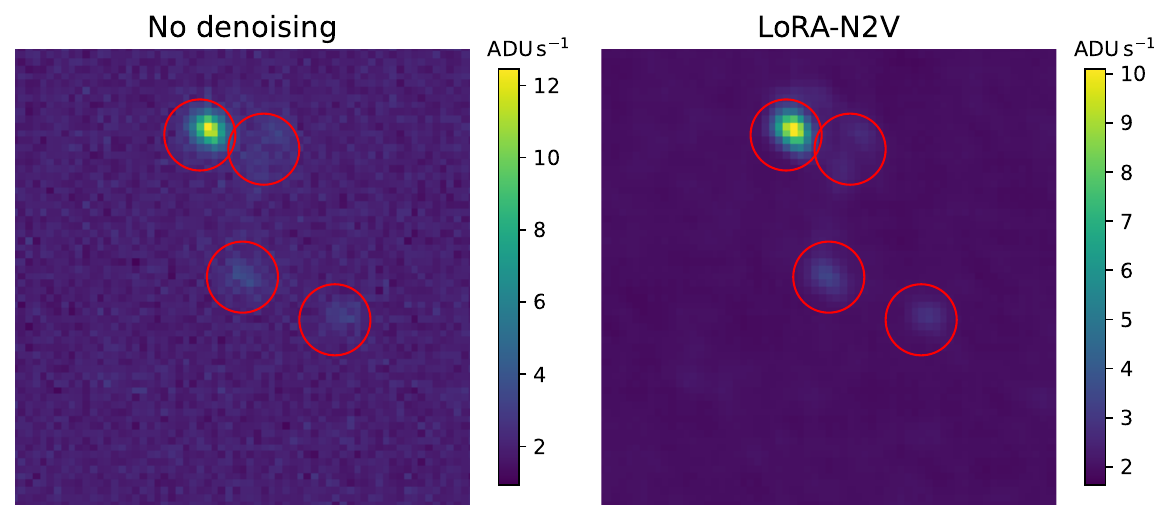}%use pdf
      \caption{Examples of PAUS stamps near the faint-end magnitude limit of the parent catalogue, shown before and after denoising. Red circles mark catalogue source positions within each stamp.}
      \label{fig:galaxieswithcircle}
\end{figure*}

Near the faint-end limit, for example at $i_{\mathrm{AB}} \approx 23$, target sources can be visually difficult to distinguish from the background noise, especially when located close to much brighter neighbouring galaxies. Figure ~\ref{fig:galaxieswithcircle} shows examples of such cases, with red circles marking the catalogue source positions. The denoised images show reduced small-scale noise, but the examples also illustrate why visual inspection alone is insufficient for assessing photometric performance near the catalogue limit.

\end{appendix}

\end{document}